\documentclass[letterpaper]{article} \usepackage{aaai2027} \usepackage[hyphens]{url} \usepackage{graphicx}   \usepackage{natbib} \usepackage{caption}  \usepackage{booktabs}
\usepackage{tabularx}
\usepackage{xurl} \newcommand{\NSnapshots}{632}
\newcommand{\NHours}{158}

\newcommand{\NEntries}{63,200}
\newcommand{\NSeries}{541}

\newcommand{\SnapshotTime}{13 November 2026, 04:24 UTC}
\newcommand{\NClientRuns}{157}

\newcommand{\NClientSeries}{461}
\newcommand{\NClientWebOverlap}{113}

\newcommand{\MinMono}{.992}
\newcommand{\NNewOverall}{3.0}
\newcommand{\MaxNewSub}{.8}
\newcommand{\MedHoursOverall}{26}

\newcommand{\RefreshHourUTC}{02:00}

\newcommand{\ShareRefreshAtModal}{.611}
\newcommand{\NRefreshDays}{7}

\newcommand{\ShareHeatChange}{.041}

\newcommand{\ShareLikeChange}{.804}
\newcommand{\ShareFavChange}{.588}

\newcommand{\RhoTenureHeat}{.783}

\newcommand{\EpMedLiveaction}{73}

\newcommand{\EpMedAIgenerated}{127}

\newcommand{\EpMedAnimated}{173}

 \newcommand{\HfeN}{2,337}
\newcommand{\HfeSeries}{443}

\newcommand{\HfeRtwo}{.923}
\newcommand{\HfeLikeFavCor}{.991}
\newcommand{\HfeWithinRtwo}{.220}
\newcommand{\HfeNoSeriesRtwo}{.190}
\newcommand{\HfeTotal}{.069}
\newcommand{\HfeTotalLo}{.049}
\newcommand{\HfeTotalHi}{.089}

\newcommand{\HfeNoSeriesLike}{.049}
\newcommand{\HfeNoSeriesLikeLo}{.026}
\newcommand{\HfeNoSeriesLikeHi}{.073}

 \newcommand{\CoverageSeries}{889}
\newcommand{\CoverageLive}{151}
\newcommand{\CoverageAI}{172}
\newcommand{\CoverageAnimated}{219}
\newcommand{\CoverageUnmatched}{348}

\newcommand{\BoardClientWebJaccard}{.084}
\newcommand{\BoardClientWebCommon}{15}

\newcommand{\RetClientComicsDay}{51.5}

\newcommand{\RetClientShortsDay}{49.0}

\newcommand{\RetWebAIDay}{92.1}

\newcommand{\RetWebAnimDay}{77.7}

\newcommand{\RetWebAllSpells}{469}

\newcommand{\RetWebAllMedian}{3}
\newcommand{\RetWebAllDay}{35.1}
\newcommand{\RetWebAllDayLo}{31.1}
\newcommand{\RetWebAllDayHi}{39.4}

\newcommand{\RetWebLiveDay}{93.6}

\newcommand{\BoardPairedHours}{146}
\newcommand{\BoardBalancedPairs}{114}
\newcommand{\BoardBalancedClient}{69.2}
\newcommand{\BoardBalancedAI}{89.4}
\newcommand{\BoardBalancedLive}{92.1}
\newcommand{\BoardBalancedAll}{83.9}
 \newcommand{\ShRows}{303}
\newcommand{\ShSeries}{22}
\newcommand{\ShAllSeries}{292}

\newcommand{\ShUp}{71}
\newcommand{\ShDown}{214}
\newcommand{\ShSingleShare}{8.037}
\newcommand{\ShSingleShareCI}{[$.617,15.457$]}
\newcommand{\ShSingleShareP}{.035}
\newcommand{\ShSingleShareQ}{.117}
\newcommand{\ShSingleLike}{6.823}
\newcommand{\ShSingleFav}{6.790}
\newcommand{\ShSingleCom}{4.983}
\newcommand{\ShJointShare}{26.420}
\newcommand{\ShJointShareCI}{[$-1.459,54.299$]}

 \newcommand{\CvN}{541}
\newcommand{\CvNLabelled}{541}

\newcommand{\CvModel}{jev-1.14.0}
\newcommand{\CvNConform}{449}
\newcommand{\CvNNot}{92}
\newcommand{\CvPctConform}{83.0}
\newcommand{\CvPctNot}{17.0}
\newcommand{\CvDiffLike}{-.257}
\newcommand{\CvLoLike}{-.539}
\newcommand{\CvHiLike}{.025}
\newcommand{\CvTLike}{-1.80}
\newcommand{\CvDfLike}{126.7}
\newcommand{\CvPLike}{= .074}

\newcommand{\CvDiffFav}{-.169}
\newcommand{\CvLoFav}{-.441}
\newcommand{\CvHiFav}{.104}
\newcommand{\CvTFav}{-1.22}
\newcommand{\CvDfFav}{126.4}
\newcommand{\CvPFav}{= .224}

\newcommand{\CvDiffComment}{.005}
\newcommand{\CvLoComment}{-.347}
\newcommand{\CvHiComment}{.356}

\newcommand{\CvPComment}{= .980}

\newcommand{\CvDiffShare}{-.206}
\newcommand{\CvLoShare}{-.615}
\newcommand{\CvHiShare}{.202}

\newcommand{\CvPShare}{= .317}
\newcommand{\CvNShare}{373}
\newcommand{\CvDiffHeat}{-.010}
\newcommand{\CvLoHeat}{-.030}
\newcommand{\CvHiHeat}{.011}

\newcommand{\CvPHeat}{= .351}

 \newcommand{\TwRuns}{157}
\newcommand{\TwAppear}{1,256}
\newcommand{\TwTopics}{95}
\newcommand{\TwWords}{309}
\newcommand{\TwTokens}{382}
\newcommand{\TwTokensW}{5,299}

\newcommand{\TwTopWordN}{21}
\newcommand{\TwTopWordNW}{204}
\newcommand{\TwSecond}{cultivation}
\newcommand{\TwSecondN}{6}
\newcommand{\TwSecondW}{male lead}
\newcommand{\TwThirdW}{comeback}
\newcommand{\TwMedianRuns}{20}
\newcommand{\TwSpells}{147}

\newcommand{\TwSlots}{8}
\newcommand{\TwSlate}{48}
\newcommand{\TwShort}{99}
\newcommand{\TwReturn}{44}

 \newcommand{\CcTagSeries}{541}
\newcommand{\CcTagTests}{143}
\newcommand{\CcTagSig}{32}
\newcommand{\CcRomanceN}{104}
\newcommand{\CcRuralOrange}{14.1}
\newcommand{\CcRuralOrangeQ}{< .001}
\newcommand{\CcPeriodOrange}{11.8}
\newcommand{\CcPeriodOrangeQ}{< .001}
\newcommand{\CcUrbanBlack}{5.2}
\newcommand{\CcUrbanBlackQ}{= .002}
\newcommand{\CcPeriodBlack}{-6.2}
\newcommand{\CcPeriodBlackQ}{= .002}
\newcommand{\CcOtherOrange}{-9.5}

\newcommand{\CcOtherBlue}{6.0}

\newcommand{\CcMeloRed}{5.6}
\newcommand{\CcMeloRedQ}{= .029}
\newcommand{\CcRomBlack}{-4.9}
\newcommand{\CcRomBlackQ}{= .025}
\newcommand{\CcRomRed}{2.5}
\newcommand{\CcRomRedQ}{= .380}
\newcommand{\CcTopATag}{rural}
\newcommand{\CcTopAColor}{orange}
\newcommand{\CcTopADiff}{14.1}
\newcommand{\CcTopAT}{8.58}
\newcommand{\CcTopADf}{128.0}
\newcommand{\CcTopAQ}{< .001}
\newcommand{\CcTopAN}{87}
\newcommand{\CcTopBTag}{period}
\newcommand{\CcTopBColor}{orange}
\newcommand{\CcTopBDiff}{11.8}
\newcommand{\CcTopBT}{6.50}
\newcommand{\CcTopBDf}{95.3}
\newcommand{\CcTopBQ}{< .001}
\newcommand{\CcTopBN}{70}
 \newcommand{\CoverPilotN}{100}
\newcommand{\CoverPilotPersonN}{97}
\newcommand{\CoverPilotTextN}{98}
\newcommand{\CoverPilotTextMedian}{20.37}
\newcommand{\CoverPilotLowerN}{71}
\newcommand{\CoverPilotAIN}{24}
\newcommand{\CoverPilotAIMedian}{22.78}
\newcommand{\CoverPilotAnimatedN}{22}
\newcommand{\CoverPilotAnimatedMedian}{19.09}
\newcommand{\CoverPilotLiveN}{26}
\newcommand{\CoverPilotLiveMedian}{18.65}
\newcommand{\CoverPilotUnknownN}{28}

 \newcommand{\SwTitles}{889}
\newcommand{\SwSeasons}{183}
\newcommand{\SwSynopses}{541}
\newcommand{\SwAppear}{92,317}
\newcommand{\SwTiTokens}{2,729}
\newcommand{\SwTiWords}{1,688}
\newcommand{\SwTiA}{rebirth}
\newcommand{\SwTiAN}{37}
\newcommand{\SwTiB}{whole family}
\newcommand{\SwTiBN}{35}
\newcommand{\SwTiC}{apocalypse}
\newcommand{\SwTiCN}{22}

\newcommand{\SwTiAW}{rebirth}
\newcommand{\SwTiAWN}{3,699}
\newcommand{\SwTiBW}{return}

\newcommand{\SwTiCW}{whole family}

\newcommand{\SwSyTokens}{21,644}
\newcommand{\SwSyWords}{9,201}
\newcommand{\SwSyA}{identity}
\newcommand{\SwSyAN}{55}
\newcommand{\SwSyB}{awakening}
\newcommand{\SwSyBN}{54}
\newcommand{\SwSyC}{mother}
\newcommand{\SwSyCN}{50}

\newcommand{\SwSyAW}{identity}

\newcommand{\SwSyBW}{mother}

\newcommand{\SwSyCW}{family}

\newcommand{\SwKinShare}{54.2}
\newcommand{\SwKinTitleShare}{11.7}
\newcommand{\SwKinTopTwenty}{7}
\newcommand{\SwKinA}{mother}
\newcommand{\SwKinB}{family}
\newcommand{\SwKinC}{husband}
 
\title{HongguoTrending: A Longitudinal, Multimodal Archive of Trending Videos and Topics on a Leading Micro-Drama Platform}
\author{Rongxin Ouyang}
\affiliations{
Department of Communications and New Media, National University of Singapore\\
rongxin@u.nus.edu
}

\renewcommand{\bibliographystyle}[1]{}
\begin{document}

\maketitle

\begin{abstract}
Micro-dramas (\textit{duanju}) are reshaping digital production and consumption, yet a lack of longitudinal data connecting series to their platform visibility persists, limiting our understanding of the emerging medium. We introduce \textit{HongguoTrending}, a longitudinal, multimodal dataset of hourly snapshots from seven trending lists of Hongguo, China's largest micro-drama platform. Collected through a two-stage asynchronous pipeline, the archive links \NEntries{} ranking entries across \NSnapshots{} web list-hours and \NClientRuns{} client hours (\CoverageSeries{} distinct series) to metadata, cover images, sampled episode videos, and audience engagement metrics. Exploratory analyses show distinct turnover rates between web and mobile trending lists, production-driven mainstream values alongside kinship-centered synopses, genre-specific cover palettes, and strong associations between user sharing and next-day rank gains. Accessible via a RESTful API, HongguoTrending supports empirical research on algorithmic gatekeeping, popularity dynamics, and multimodal digital culture.
\end{abstract}

\begin{links}
\link{Dataset}{https://micro-drama.rongxin.me/}
\par\textbf{Code} --- To be open-sourced later.
\end{links}

\begin{figure*}[!t]
\centering
\includegraphics[width=0.8\textwidth]{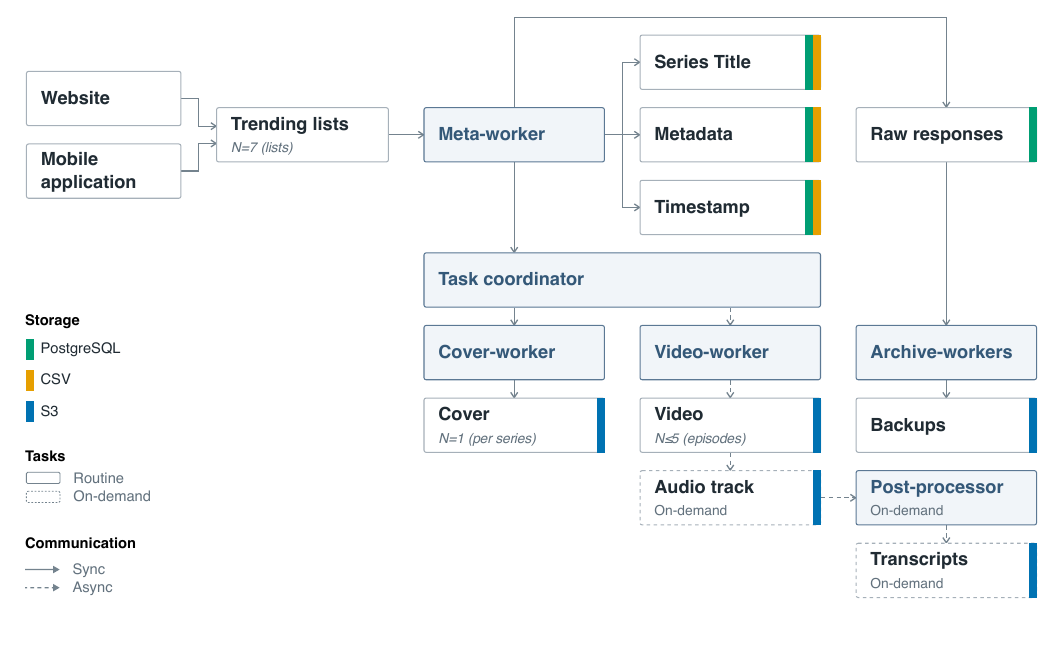}
\caption{Data modalities, collection, and storage.}
\label{fig:pipeline}
\end{figure*}

\section{Introduction}
Micro-dramas\footnote{\textit{Duanju} in Mandarin, literally ``short drama''.} are reshaping digital production and consumption \cite{liu2026producing,cao2026audience}. This new form of storytelling consists of high-volume, serialized narratives made up of brief vertical episodes that have grown popular worldwide \cite{liu2026producing,ita2025microdrama,hanney2026new}. In 2024, China's micro-drama market generated 50.4 billion yuan in revenue, surpassing the national film box office for the first time \cite{caixin2024boxoffice}. Whereas early growth relied on pay-per-episode mini-programs, the sector is now led by free, ad-supported applications \cite{caixin2024boxoffice}. ByteDance's Hongguo, the largest of these platforms in China, reached 430 million monthly active users as of September 2026 \cite{questmobile2026autumn}. Beyond China, international apps such as ReelShort and PineDrama have extended this model abroad \cite{tang2025reelshort,ita2025microdrama,hanney2026new}, building a \$4 billion overseas market \cite{dataeye2025haiwai}. Combining rapid serial production with algorithmic distribution, micro-drama platforms provide an important empirical setting for studying platform attention, creative labor, and transnational media flows \cite{liu2026producing,tang2025reelshort,hanney2026new,dataeye2025haiwai}.

However, a lack of longitudinal data connecting micro-drama series to their changing platform visibility limits our understanding of micro-dramas. Early qualitative studies illuminate production workflows and cross-border circulation \cite{tang2025reelshort,liu2026producing,hanney2026new}. Existing datasets available for quantitative research are also limited. The first branch focuses on the quality of video content \cite{liu2026bridging}, while the second concerns the understanding of micro-dramas \cite{zhong2026storytr,qin2026sparse}. However, neither of these branches has the capability for longitudinal analysis in the context of platform visibility, which is theoretically important \cite{yang2020importance,gillespie2014relevance}.

We address this gap with \textit{HongguoTrending}, a longitudinal, multimodal dataset collected on an hourly schedule from seven trending lists across Hongguo's web and mobile interfaces. Collection attempts to retrieve all reported trending-list positions and records observation status. Series records link to metadata, cover images, video episodes, and audience engagement metrics. In the remainder of this paper, we describe the data collection and structure (Figure~\ref{fig:pipeline}), report exploratory analyses of list turnover, textual themes, cover design, and audience engagement, and briefly discuss future research opportunities.

\section{Research Gaps}

\paragraph{Lack of longitudinal archives.}
Trending lists are a form of platform gatekeeping \cite{yang2020importance}, and existing archives already preserve longitudinal records of YouTube trending videos and real-time Google search trends \cite{goncalves2026global,urman2026googletrends}. However, to the best of our knowledge, no longitudinal dataset preserves such records for micro-dramas. Our dataset extends this archival approach to the micro-drama industry.

\paragraph{Limited connection between engagement and visibility.}
Although some short-drama datasets include audience feedback metrics, these metrics are either static snapshots \cite{liu2026bridging} or do not reflect users' actual behavior on the platform \cite{wang2025dramaad}. We link ratings and audience engagement metrics to series and trending-list records, thus allowing longitudinal analyses of such behaviors.

\paragraph{Missing connection between visibility and multimodal content.}
Although the aforementioned datasets contain multimodal content, they lack the ability to connect this content with platform visibility \cite{zhong2026storytr,qin2026sparse,liu2026bridging}. We address this limitation by linking trending lists to the available modalities to enrich analyses of how content characteristics relate to algorithmic attention.

\section{Dataset}
\subsection{Collection}
We collect seven trending lists across Hongguo's web interface and mobile application. On the web interface, we track four lists, namely overall, live-action (human actors), AI-generated (realistic AI-rendered series), and animated (\textit{manju}). On the mobile client, we use the open-source \texttt{openhongguo} SDK \cite{openhongguo} to retrieve three additional lists, namely short dramas (combining live-action and AI-generated titles), animated micro-dramas, and an editorially curated trending-topic list. Together, these seven streams capture platform-curated attention across both interfaces.

Our collection uses a two-stage pipeline consisting of a metadata worker and an asset scheduler (Figure~\ref{fig:pipeline}). In each hourly run, the metadata worker extracts series attributes (title, series ID, regulatory filing number, episode count, search score, cast, and genre tags) and writes them to PostgreSQL and CSV files. The scheduler then dispatches three asynchronous workers to fetch cover images, episode videos, and archival snapshots, storing multimodal objects on a RustFS-backed, S3-compatible server \cite{rustfs}. Collection schedules and coverage rules are detailed in Table~\ref{tab:collection-protocol} (Appendix~\ref{app:collection-protocol}).

\subsection{Structure and access}
As of \SnapshotTime{}, the archive spans \NHours{} consecutive hours of web-list data and \NClientRuns{} hours of client-side data, covering \CoverageSeries{} distinct series (\CoverageLive{} live-action, \CoverageAI{} AI-generated, \CoverageAnimated{} animated, and \CoverageUnmatched{} without a matched website category; Table~\ref{tab:components}).

We provide a RESTful API for programmatic access to list snapshots, series metadata, engagement histories, and cover assets (request and response examples in Appendix~\ref{app:api}; browsing interface in Appendix~\ref{app:frontend}). Public browsing is open, while, to manage server load, bulk API queries require a researcher key available from the authors.
\begin{table*}[t]
\centering
\small
\caption{\textit{Components of the dataset}}
\label{tab:components}
\begin{tabularx}{\textwidth}{@{}>{\raggedright\arraybackslash}p{3.2cm}>{\raggedright\arraybackslash}p{3.4cm}r>{\raggedright\arraybackslash}X@{}}
\toprule
Component & Unit & Count & Principal fields \\
\midrule
Series & Distinct series ID & 906 & all tables; 889 on ranking lists, website detail metadata for 541 \\
Series videos & Episode file & 2,689 & 541 series; episode number, duration, size, SHA-256, object key (24.82 GB) \\
Series unique casts & Cast member & 1,244 & name, role, avatar URL, linked series (151 series list cast) \\
Series covers & Image file & 1,007 & SHA-256, format, size, linked series and capture time \\
Rank items & List $\times$ hour $\times$ position & 92,317 & website 63,200, client 29,117; rank, series ID, heat or search score \\
Covered ranking hours & Hour & 158 & website 158, client 157 scheduled hours \\
Covered engagement hours & Hour & 158 & 63,200 series-hour readings of heat, rating, likes, favorites, comments \\
\bottomrule
\end{tabularx}
\par\smallskip\raggedright\footnotesize\textit{Note.} Full database snapshot: 13 November 2026, 04:24 UTC. Website and client hours overlap.
\end{table*}
 
\section{Analyses}
Using the initial archive snapshot (through \SnapshotTime{}), we examine four aspects of the platform, namely (a) composition and turnover of trending \textit{lists}, (b) textual themes in \textit{series} and \textit{topics}, (c) visual properties of series \textit{covers}, and (d) longitudinal trajectories of \textit{audience engagement}. Notably, these preliminary analyses will be updated later as more data are collected and the archive evolves.

\subsection{Overview of trending lists}
\begin{figure*}[!t]
\centering
\includegraphics[width=0.9\textwidth]{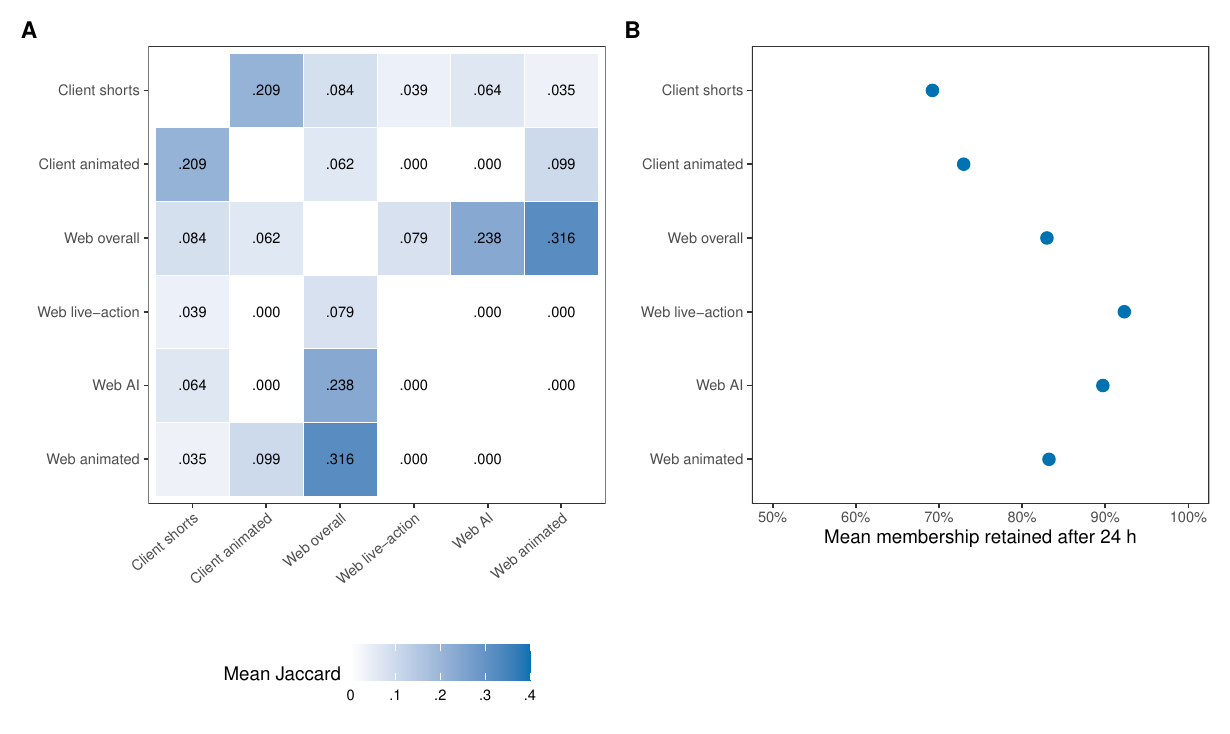}
\caption{Client and website lists differ in membership and turnover. Panel (A) shows overlap as mean same-hour Jaccard similarity. Panel (B) shows turnover as the fraction of series remaining listed after 24 hours.}
\label{fig:boardcomparison}
\end{figure*}

\paragraph{Similarities and differences across lists.}
First, we compare series overlap and turnover across the client and website lists (Table~\ref{tab:boards}). The two interfaces surface largely separate catalogs at different speeds. One, cross-interface overlap is minimal, since across \BoardPairedHours{} paired hours the client short-drama list and the website overall list share a median of \BoardClientWebCommon{} titles, with a mean Jaccard similarity of \BoardClientWebJaccard{} (Figure~\ref{fig:boardcomparison}A). Two, the client list turns over much faster, as only \BoardBalancedClient\% of client series remain listed across \BoardBalancedPairs{} 24-hour intervals, compared with \BoardBalancedAll\% on the website overall list, \BoardBalancedAI\% on the AI list, and \BoardBalancedLive\% on the live-action list (Figure~\ref{fig:boardcomparison}B). Web and mobile trending lists thus represent distinct curation channels rather than mirrors of a single ranking (Appendix~\ref{app:descriptive}).

\begin{table*}[t]
\centering
\small
\caption{\textit{Coverage, turnover, and series attributes by trending list}}
\label{tab:boards}
\setlength{\tabcolsep}{2pt}
\begin{tabular*}{\textwidth}{@{\extracolsep{\fill}}lrrrrrrrrrr@{}}
\toprule
 & \multicolumn{3}{c}{Coverage} & \multicolumn{3}{c}{Turnover} & \multicolumn{4}{c}{Series attributes (median per series)} \\
\cmidrule(lr){2-4}\cmidrule(lr){5-7}\cmidrule(lr){8-11}
 & & & & Hours & New per & Top-10 & Rating & Heat & Likes & \\
Trending list & Snapshots & Entries & Series & on list & hour & kept & \textit{M} (\textit{SD}) & (M) & (K) & Episodes \\
\midrule
Overall & 158 & 15,800 & 300 & 26 & 2.987 & .967 & 8.302 (.816) & 53.293 & 697.500 & 102 \\
Live-action & 158 & 15,800 & 151 & 123 & .350 & .995 & 8.247 (.644) & 46.472 & 353.000 & 73 \\
AI-generated & 158 & 15,800 & 172 & 90 & .484 & .991 & 8.217 (.710) & 50.104 & 630.500 & 127 \\
Animated & 158 & 15,800 & 219 & 51 & .758 & .990 & 8.214 (.964) & 49.886 & 674.000 & 173 \\
\bottomrule
\end{tabular*}
\par\smallskip\raggedright\footnotesize\textit{Note.} Hours on list: median observed snapshots per series. New per hour and Top-10 kept: mean additions and retention from the preceding hour. Attributes use per-series medians; rating summaries exclude missing ratings. Heat in millions; likes in thousands.
\end{table*}
 
\begin{figure*}[!t]
\centering
\includegraphics[width=0.9\textwidth]{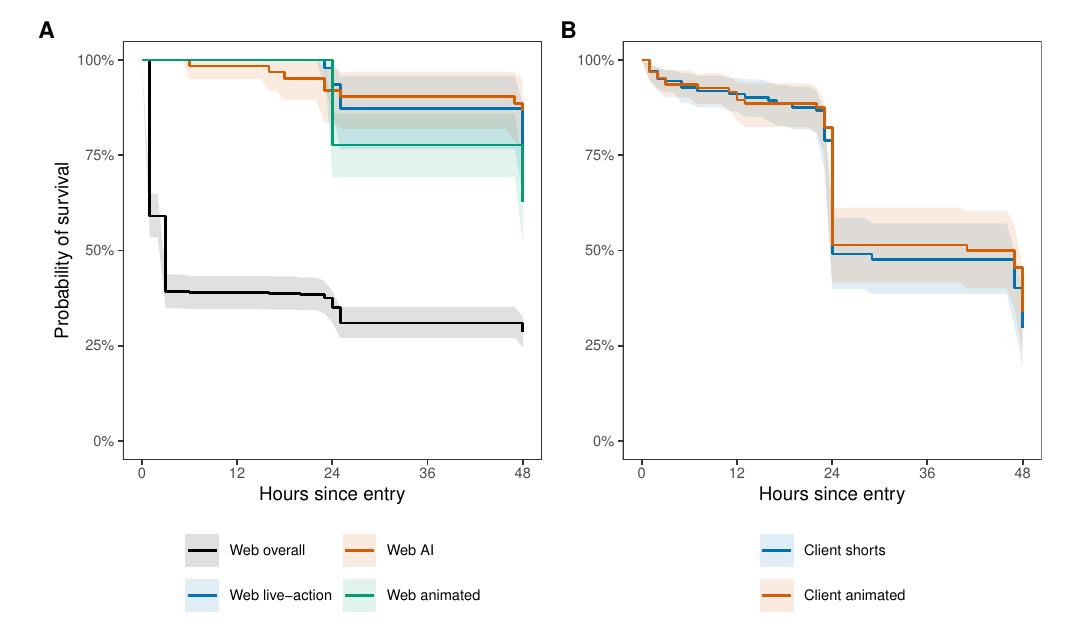}
\caption{Newly listed series often leave within a day, at a pace set by each list. (A) Website lists. (B) Client lists. Kaplan--Meier probability that a newly listed series is still listed, with bands showing 95\% series-bootstrap intervals (1,000 replicates). }
\label{fig:decaynew}
\end{figure*}

\paragraph{Rapid turnover of trending series and topics.}
Second, we measure how quickly new entrants exit the trending lists. One, newly listed series often leave within a day, at a pace specific to each list. On the website overall list, half of \RetWebAllSpells{} entry spells end within \RetWebAllMedian{} hours and \RetWebAllDay\% last 24 hours ([\RetWebAllDayLo, \RetWebAllDayHi]\%; Figure~\ref{fig:decaynew}A) across entry ranks. Category lists are far more stable, with \RetWebLiveDay\% of live-action, \RetWebAIDay\% of AI, and \RetWebAnimDay\% of animated entrants remaining after 24 hours. On the client, \RetClientShortsDay\% of short-drama and \RetClientComicsDay\% of animated entrants last 24 hours, with both survival curves dropping sharply at the daily refresh (Figure~\ref{fig:decaynew}B). For comparison, Weibo trending topics have a median survival of 6.28 hours \cite{liu2026unveiling}.

Two, client trending topics update in daily batches rather than continuously. Although the topic module displays \TwSlots{} phrases at a time, only \TwTopics{} distinct topics appear across \TwRuns{} hours. Of \TwSpells{} topic spells, \TwSlate{} run for 20--24 hours as a daily slate and \TwShort{} last five hours or less, while all replacements occur within the same four-hour daily window and \TwReturn{} topics reappear after leaving. Editorial topics therefore follow a scheduled daily rotation, unlike the hourly churn of the website overall list (Appendices~\ref{app:topicwords} and~\ref{app:serieswords}).

\subsection{Textual descriptions of series and topics}
\begin{figure*}[!t]
\centering
\includegraphics[width=0.9\textwidth]{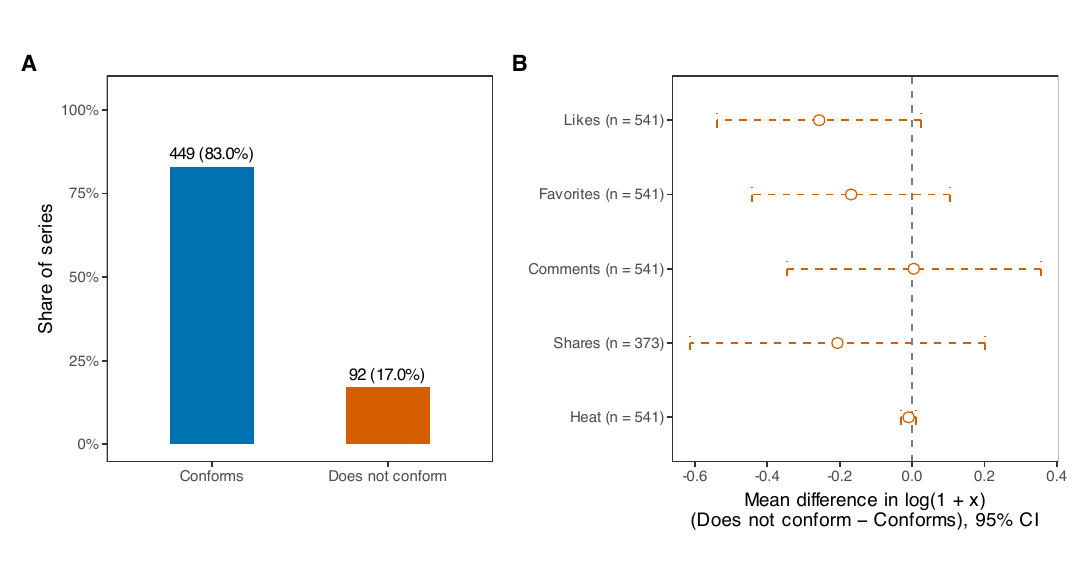}
\caption{Core socialist values prevail among listed series, independent of engagement. (A) Share and count of series labeled as conforming or not conforming to core socialist values, based on title and synopsis. (B) Mean difference in $\log(1+x)$ of each series' latest website reading (not conforming minus conforming), with Welch 95\% confidence intervals, where dashed lines and open points mark intervals that include zero. }
\label{fig:corevalues}
\end{figure*}

\paragraph{Prevalent mainstream values not driven by engagement.}
Third, we examine the value orientation of listed series. Core socialist values prevail across the catalog but do not covary with audience engagement. One, value conformity is widespread, as a closed-choice language model given each series' title and synopsis labels \CvNConform{} of \CvN{} website-listed series (\CvPctConform\%) as conforming and \CvNNot{} (\CvPctNot\%) as non-conforming (Figure~\ref{fig:corevalues}A; Appendix~\ref{app:corevalues}). Two, conforming and non-conforming series attract similar levels of likes, favorites, comments, shares, and heat, with all confidence intervals spanning zero or showing negligible differences (Figure~\ref{fig:corevalues}B). This pattern indicates that mainstream value alignment on the trending lists stems primarily from production supply rather than selective audience demand.

\begin{figure*}[!t]
\centering
\includegraphics[width=0.9\textwidth]{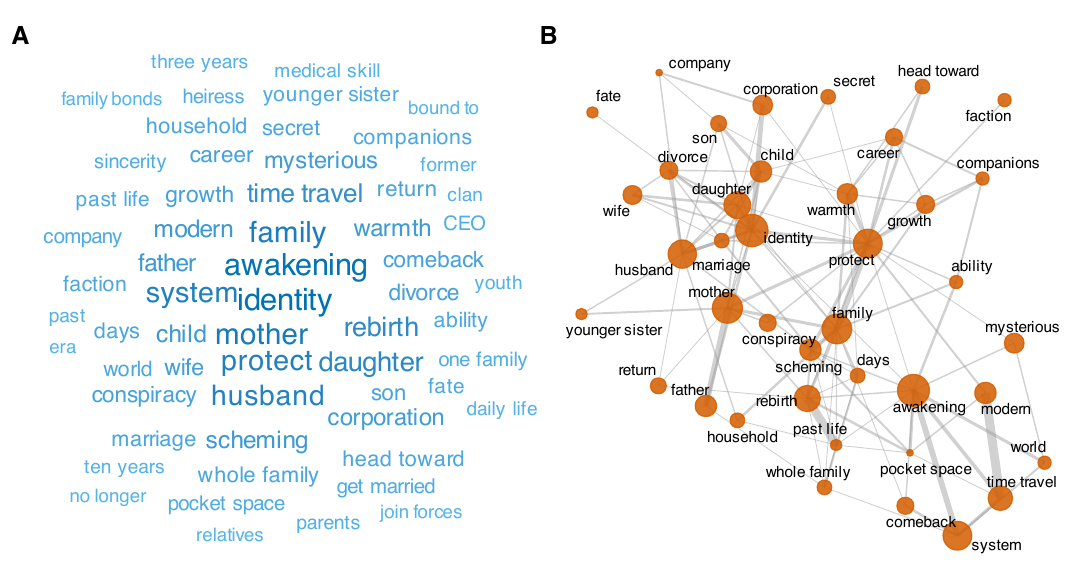}
\caption{Kinship dominates series synopses. (A) Word cloud of the 60 most frequent words, in which the size is proportional to frequency. Text is segmented with Jieba \cite{sun2012jieba}. (B) Co-occurrence network of the most frequent words ($40$) and the strongest links ($120$), in which the node size shows frequency and edge width the number of synopses in which two words co-occur.}
\label{fig:synopsisdedup}
\end{figure*}

\paragraph{Kinship at the center of series synopses.}
Fourth, we analyze the vocabulary of series synopses. Kinship dominates how series describe themselves. One, family terms are pervasive, with \SwKinShare\% of the \SwSynopses{} website synopses naming at least one kin relation or the family as a unit, against \SwKinTitleShare\% of series titles. Two, they dominate the vocabulary, accounting for \SwKinTopTwenty{} of the 20 most frequent synopsis words, led by \textit{\SwKinA}, \textit{\SwKinB}, and \textit{\SwKinC} (Figure~\ref{fig:synopsisdedup}A). Three, they anchor the plot structure, as mother, husband, daughter, and family form the dense core of the co-occurrence network, linked to identity, awakening, and protection (Figure~\ref{fig:synopsisdedup}B). This attention to kinship likely reflects the cultural emphasis on close familial and interpersonal relationships in Fei's work (\citeauthor{fei1992soil} \citeyear{fei1992soil}; details in Appendix~\ref{app:serieswords}).

\subsection{Visual properties of covers}
\begin{figure*}[htbp]
\centering
\includegraphics[width=.9\textwidth]{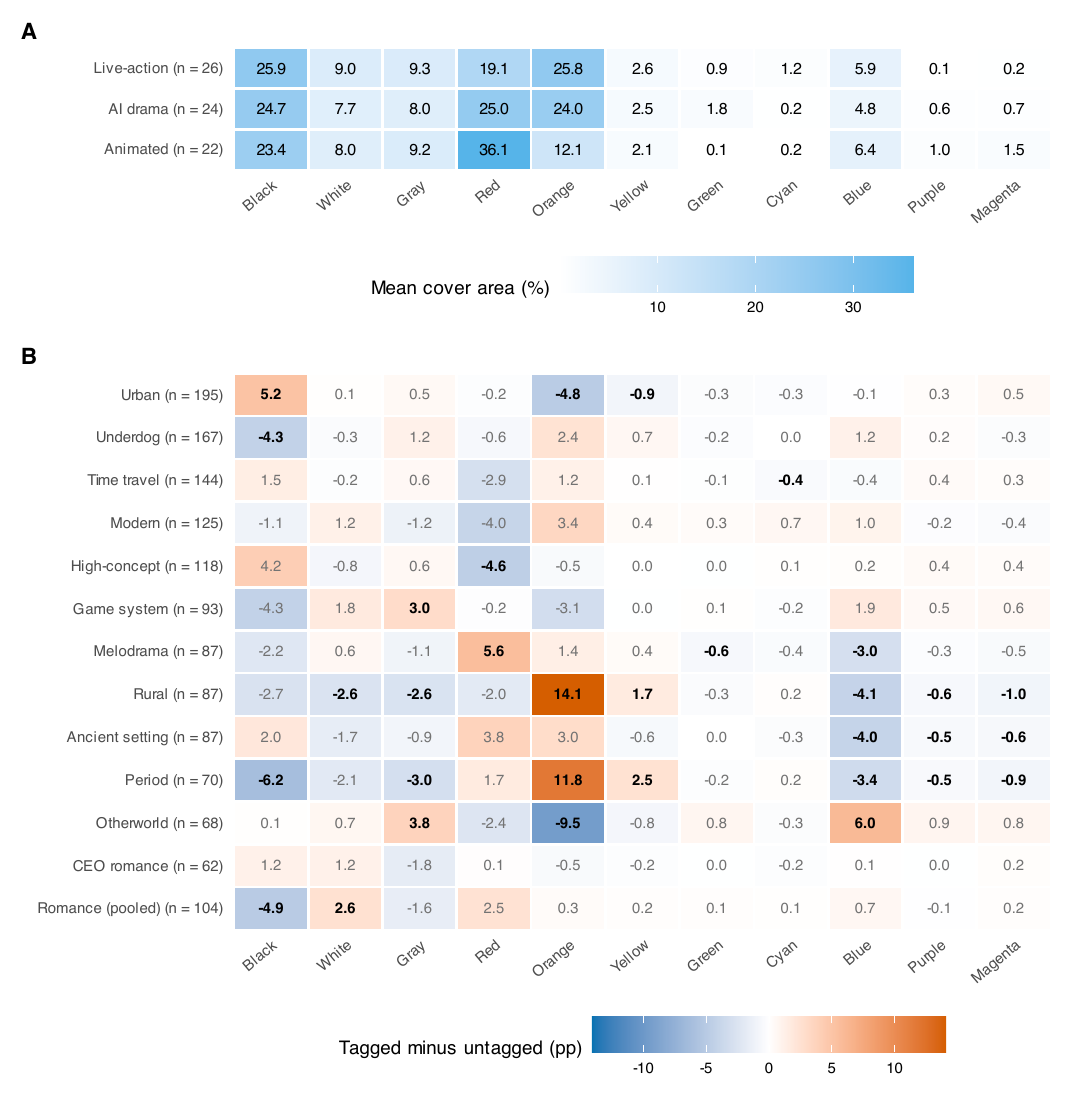}
\caption{Cover colors signal a series' setting more than its production format. (A) Mean whole-cover color shares by official category across 72 client series. (B) Difference in mean color share (percentage points) between website series with and without each tag across \CcTagSeries{} series, where orange marks colors more common among tagged series and blue less common ones. Bold cells are BH-corrected $q<.05$ across \CcTagTests{} cells.}
\label{fig:covercolor}
\end{figure*}

\paragraph{Genre-specific color palettes in series covers.}
Fifth, we compare cover colors across genre tags. Cover colors signal a series' setting more than its production format. One, warm palettes mark nostalgic settings, since among \CcTagSeries{} website series, covers tagged \textit{\CcTopATag} (\textit{n}~=~\CcTopAN) carry \CcTopADiff{} percentage points more \CcTopAColor{} area than other covers, $t(\CcTopADf)=\CcTopAT$, $q \CcTopAQ$, and covers tagged \textit{\CcTopBTag} (\textit{n}~=~\CcTopBN) carry \CcTopBDiff{} points more \CcTopBColor, $t(\CcTopBDf)=\CcTopBT$, $q \CcTopBQ$ (Figure~\ref{fig:covercolor}B). Two, other settings have their own palettes, as urban covers carry more black and otherworld covers more blue, and \CcTagSig{} of \CcTagTests{} tag-color contrasts hold after BH correction. Three, production format matters less, because although animated client covers show more red than live-action or AI covers, the overall category difference is not significant (Figure~\ref{fig:covercolor}A; Appendix~\ref{app:covercolor}). These visual patterns suggest category-specific production choices that merit further exploration.

\begin{figure*}[!t]
\centering
\includegraphics[width=\textwidth]{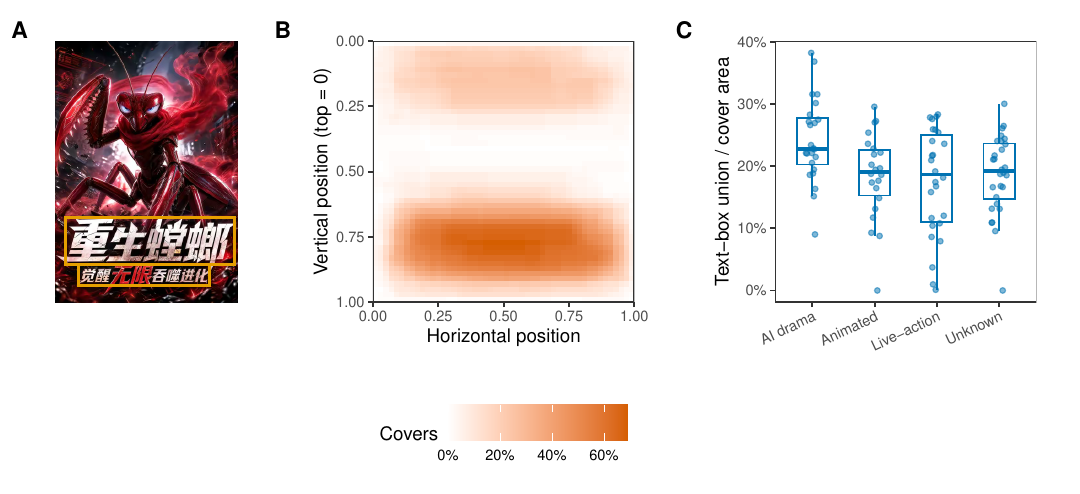}
\caption{Cover-text layout measured with Apple Vision OCR in all panels ($N=\CoverPilotN{}$). (A) Detected text boxes on \textit{Reborn as a Mantis} (shortened title). (B) Fraction of covers with text boxes at each normalized position, with zero at the top. (C) Text-box union as a share of cover area, by category. In this descriptive pilot, unknown denotes an unmatched category.}
\label{fig:covercomposition}
\end{figure*}

\paragraph{Consistent lower-half text layout on covers.}
Sixth, we examine the spatial layout of cover elements in a pilot sample ($N=\CoverPilotN{}$). Posters follow a consistent composition that pairs character imagery with lower-frame title overlays. One, human figures and title text appear on nearly every poster, as automated detection identifies people in \CoverPilotPersonN{} covers and text in \CoverPilotTextN{}. Two, typography occupies a large footprint in the bottom half, as detected text boxes cover a median of \CoverPilotTextMedian\% of image area, with their combined centroid falling in the lower half on \CoverPilotLowerN{} of the \CoverPilotTextN{} text-detected covers (Figure~\ref{fig:covercomposition}). This recurring layout leaves the upper frame for character portraits while anchoring titles at the bottom across production categories (Appendix~\ref{app:covercomposition}).

\subsection{Longitudinal trajectories of audience engagement}
\begin{figure*}[!t]
\centering
\includegraphics[width=0.9\textwidth]{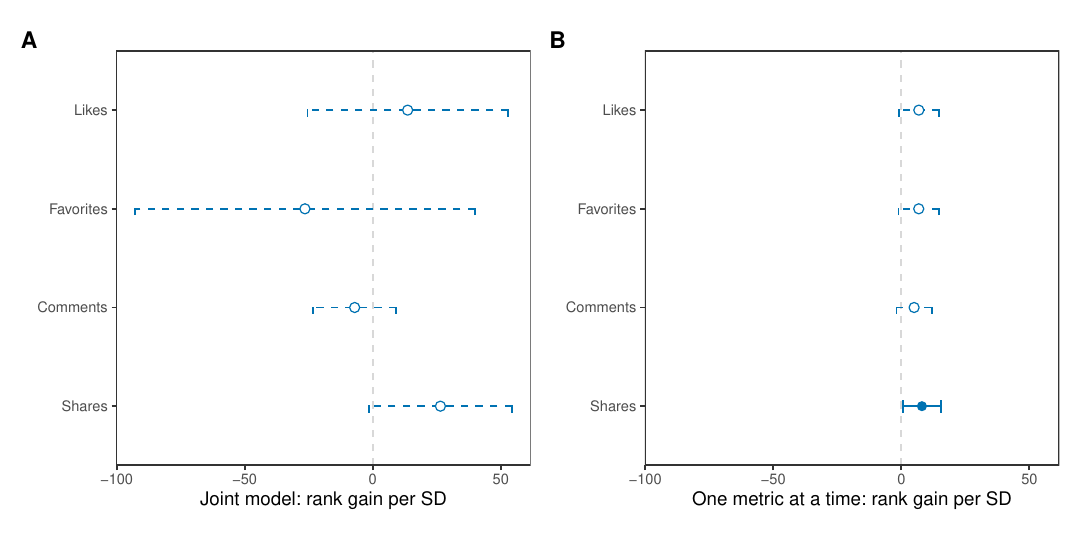}
\caption{Sharing is the strongest single correlate of next-day rank gain. Engagement-metric coefficients on a common sample, from (A) a joint model and (B) one metric at a time, with a shared x-axis. The dashed grey line marks zero, and bars show unadjusted 95\% clustered intervals. Solid lines and filled points mark intervals that exclude zero, whereas dashed lines and open points mark intervals that include zero.}
\label{fig:sharenew}
\end{figure*}

\paragraph{Sharing as a strong signal of trending series.}
Seventh, we test which engagement metrics predict client rank changes. Sharing is the strongest single correlate of next-day rank gain. One, in single-metric regressions across \ShRows{} transitions from \ShSeries{} of \ShAllSeries{} titles (\ShUp{} gains, \ShDown{} losses), a one-SD increase in standardized $\log(1+x)$ shares predicts a gain of \ShSingleShare{} client positions over 24 hours (95\% CI \ShSingleShareCI{}; $p=\ShSingleShareP$), whereas confidence intervals for likes (\ShSingleLike{}), favorites (\ShSingleFav{}), and comments (\ShSingleCom{}) include zero (Figure~\ref{fig:sharenew}B). Two, the estimate loses precision under multiple-testing adjustment ($q=\ShSingleShareQ$) and in the joint model including all metrics simultaneously ($\beta=\ShJointShare$, CI \ShJointShareCI{}; Figure~\ref{fig:sharenew}A). Controlling for current rank, its square, and snapshot fixed effects, outward sharing tracks subsequent upward movement more closely than passive approval (Appendix~\ref{app:robustness}).

\paragraph{Weak within-series sensitivity of website heat to engagement.}
Finally, we relate the proprietary website heat score to public engagement counts. Website heat rises with cumulative engagement, but primarily reflects persistent cross-series differences. With series fixed effects $\alpha_i$ and day fixed effects $\lambda_d$, the estimated relationship is
\begin{equation}
 \widehat{\log H}_{id}=\hat\alpha_i+\hat\lambda_d+\HfeTotal\,\log\left(1+L_{id}+F_{id}+C_{id}\right),
\end{equation}
where $H_{id}$ is the heat of series $i$ on day $d$ after the daily refresh, and $L$, $F$, and $C$ denote likes, favorites, and comments (\textit{N} = \HfeN{} series-days across \HfeSeries{} series; 95\% CI [\HfeTotalLo{}, \HfeTotalHi{}]; Table~\ref{tab:heatfe}). One, within a series, a 1\% increase in combined engagement corresponds to only a \HfeTotal\% rise in heat, while likes and favorites co-move tightly ($r=\HfeLikeFavCor{}$). Two, between-series differences dominate variance, since adding series fixed effects raises $R^2$ from \HfeNoSeriesRtwo{} to \HfeRtwo{}, whereas within-series engagement explains \HfeWithinRtwo{} of within-series variation. Daily heat scores thus act largely as slow-moving baseline tiers rather than direct tallies of public interactions.
\begin{table*}[!t]
\centering\small
\caption{Website heat rises with engagement, but weakly, in two-way fixed-effects models of log website heat.}
\label{tab:heatfe}
\begin{tabular}{l*{4}{>{\centering\arraybackslash}p{2.3cm}}}
\hline
 & (1) & (2) & (3) & (4) \\
 & Total & Components & Hourly & No series FE \\
\hline
Total (likes + favorites + comments) & .069$^{***}$ &  &  &  \\
 & (.010) &  &  &  \\
Likes &  & $-$.360$^{***}$ & $-$.326$^{***}$ & .049$^{***}$ \\
 &  & (.072) & (.072) & (.012) \\
Favorites &  & .456$^{***}$ & .417$^{***}$ & $-$.027$^{*}$ \\
 &  & (.075) & (.074) & (.013) \\
Comments &  & .005 & .005 & .010$^{*}$ \\
 &  & (.009) & (.007) & (.004) \\
\hline
Series FE & Yes & Yes & Yes & No \\
Time FE & Day & Day & Hour & Day \\
$N$ & 2,337 & 2,337 & 47,043 & 2,435 \\
Series & 443 & 443 & 541 & 541 \\
$R^2$ & .911 & .923 & .935 & .190 \\
Adjusted $R^2$ & .890 & .904 & .934 & .187 \\
Within $R^2$ & .107 & .220 & .173 & .190 \\
\hline
\multicolumn{5}{p{.95\linewidth}}{\footnotesize \textit{Note.} Outcome: log website heat; predictors are log(1 + count). Columns 1, 2, and 4 use one reading per series and list-day, taken after the daily heat refresh; column 3 uses every series-hour reading. Series-clustered standard errors in parentheses. $^{*}p<.05$, $^{**}p<.01$, $^{***}p<.001$.} \\
\end{tabular}
 \end{table*}

\section{Research Potential}
HongguoTrending supports at least four directions in computational communication and platform studies \cite{nieborg2018platformization,peng2024automated}. First, it enables longitudinal analysis of micro-drama supply and reception, extending interview-based studies of creative labor and audience feedback \cite{liu2026producing,cao2026audience} as well as research on how commercial entertainment intersects with mainstream discourse on Chinese video platforms \cite{chen2021positiveenergy,lu2025decentralized}.

Second, hourly ranking logs allow researchers to audit algorithmic curation by tracing how trending-list placement relates to subsequent engagement, complementing field and quasi-experimental studies of algorithmic amplification \cite{huszar2022amplification,guess2023feeds,yang2020importance}.

Third, linking text, cover imagery, and video streams supports multimodal analyses of how visual presentation relates to audience response \cite{sharma2024visual,xue2026sensation,kayal2025popularity,peng2024automated}. For example, researchers can test whether visual styles associated with AI-generated imagery \cite{liu2026seeing} appear in micro-dramas and how thematic framing relates to online discussion and polarization \cite{mukerjee2025antecedents,guo2025civilizing,yu2024topic}.

Fourth, the archive's panel structure supports quasi-experimental evaluations of platform or policy changes when clear interventions exist \cite{baker2025did,abadie2021using,athey2017state}, such as measuring how interface updates shift engagement dynamics \cite{guo2025civilizing,yu2024topic,athey2021impact}.

\section{Limitations}
This study has four main limitations. First, querying the platform via a single anonymous profile omits user-level personalization, though trending lists are platform-wide rankings where personalization is minimal. Second, the initial observation window does not capture seasonal or long-term cycles. Third, all regression results are correlational, since the heat models estimate within-series associations with collinear engagement metrics rather than recovering a proprietary formula, sharing data cover only the final collection days, trending-list prominence may endogenously induce sharing, and value labels rely on language-model classifications of brief synopses. Fourth, data from a single domestic Chinese platform may not generalize to international apps such as ReelShort with different monetization models \cite{tang2025reelshort}.

\section{Ethical Considerations}
The archive collects only public platform aggregates and series metadata, excluding personal data from user accounts, comments, viewing histories, and device identifiers. Scrapers use bounded retries and exponential backoff to minimize server load. Because copyrights remain with the original rights holders, raw video files and transcripts are excluded from public redistribution and restricted to authorized academic use under controlled access.

\providecommand{\noopsort}[1]{}

\clearpage
\appendix
\section*{Appendix}
\noindent The appendix follows the order of the main analyses. Appendix~\ref{app:data} documents data collection and access, including the collection protocol, the multimodal inventory, and the browsing interface. Appendix~\ref{app:descriptive} adds descriptive results on the trending lists, and Appendix~\ref{app:text} reports the vocabulary of trending topics, series titles, and synopses together with the core-values labeling method. Appendix~\ref{app:covers} details cover colors and cover-text layout, Appendix~\ref{app:robustness} reports the heat specifications and correlations among engagement metrics, and Appendix~\ref{app:api} gives example API requests and responses.

\setcounter{figure}{0}
\renewcommand{\thefigure}{A\arabic{figure}}
\setcounter{table}{0}
\renewcommand{\thetable}{A\arabic{table}}
\section{Data Collection and Access}
\label{app:data}
\subsection{Collection Timing and Protocols}
\label{app:collection-protocol}
Table~\ref{tab:collection-protocol} summarizes the acquisition schedule and earliest retained record for each data component. Because network latency and pagination can delay individual runs, downstream analyses rely on recorded fetch timestamps and completeness flags rather than nominal hourly targets.

\begin{table*}[t]
\centering
\small
\caption{Acquisition timing and methods by data component.}
\label{tab:collection-protocol}
\setlength{\tabcolsep}{4pt}
\renewcommand{\arraystretch}{1.13}
\begin{tabularx}{\textwidth}{@{}>{\raggedright\arraybackslash}p{.09\textwidth}>{\raggedright\arraybackslash}p{.17\textwidth}>{\raggedright\arraybackslash}p{.10\textwidth}>{\raggedright\arraybackslash}p{.16\textwidth}>{\raggedright\arraybackslash}X@{}}
\toprule
 & Component & Start (UTC) & Interval & Explanation \\
\midrule
Website & Ranking lists & 2026-11-06\newline 15:12 & Hourly & All pages of the overall, live-action, AI, and animated lists. \\
\addlinespace[3pt]
\midrule
Client & Topic lists & 2026-11-06\newline 16:53 & Hourly & Search-page topic module, via the openhongguo SDK. \\
\addlinespace[3pt]
 & Video lists & 2026-11-06\newline 18:23 & Hourly & Search-page trending module: series, positions, and search scores. \\
\addlinespace[3pt]
 & Animated lists & 2026-11-06\newline 17:20 & Hourly & All pages of both lists, via the SDK. \\
\addlinespace[3pt]
\midrule
Engagement & Website & 2026-11-06\newline 15:12 & After hourly lists & Detail page of each listed series: engagement metrics and static metadata. \\
\addlinespace[3pt]
 & Client & 2026-11-13\newline 06:49 & After hourly lists & Series heat, plays, follows, likes, and episode-1 comments; not in the analyzed export. \\
\addlinespace[3pt]
\midrule
Multimodal & Website covers & 2026-11-06\newline 18:42 & Event-driven & Cover images from list and detail records, deduplicated. \\
\addlinespace[3pt]
 & Client covers & 2026-11-06\newline 17:30 & Event-driven & Cover images of client-listed series, deduplicated. \\
\addlinespace[3pt]
 & Episode videos & 2026-11-11\newline 17:06 & Event-driven & First five episodes per series, via openhongguo. \\
\addlinespace[3pt]
 & Audio and transcripts & On demand & On demand & Embedded in stored videos. \\
\addlinespace[3pt]
\bottomrule
\end{tabularx}
\vspace{3pt}
\begin{minipage}{\textwidth}\footnotesize
\textit{Note.} Start times are reported to the minute without rounding. Aside from the client-detail extension, earliest records refer to the database export initiated on 13 November 2026 at 04:24:53 UTC, with a latest scheduled list-hour of 04:00 UTC, and subsequent client-detail observations fall outside this analytical snapshot. Collection intervals indicate scraper schedules rather than platform refresh intervals, and S3/database backups store existing observations.
\end{minipage}
\end{table*}
 \subsection{Multimodal Coverage}
The archive pairs cover images with an expanding collection of sampled episode videos (Figure~\ref{fig:multimodal}). In the 13 November 2026 database snapshot, the analytical export includes 292 hash-verified covers alongside 2,689 downloaded episodes across 541 series (24.82 GB in total, with a median duration of 101.6 seconds). Because video downloading began later than list tracking, episode coverage varies across series, and audio extraction and transcription are scheduled for subsequent releases.

\begin{figure*}[t]
\centering
\small
\begin{tabularx}{\textwidth}{@{}>{\raggedright\arraybackslash}p{0.10\textwidth}>{\raggedright\arraybackslash}p{0.25\textwidth}>{\raggedright\arraybackslash}p{0.29\textwidth}>{\raggedright\arraybackslash}X@{}}
\toprule
Modality & Coverage & Dimensions / duration & Representation \\
\midrule
Cover & 292 series & Width 400--400 px; height 534--600 px & Earliest observed cover per series \\
Video & 2,689 episodes / 541 series & Median 101.6 s; 30--398.9 s & MP4; 24.82 GB \\
Audio & Not separately inventoried & Audio-track statistics not measured & No separate audio / transcript release \\
\bottomrule
\end{tabularx}
 \caption{Multimodal inventory in the full database snapshot of 13 November 2026, 04:24 UTC. Video collection is ongoing, and file sizes and durations reflect recorded metadata.}
\label{fig:multimodal}
\end{figure*}
 
\subsection{Frontend Demonstration}
\label{app:frontend}
\begin{figure}[htbp]
\centering
\includegraphics[width=\linewidth,height=.78\textheight,keepaspectratio]{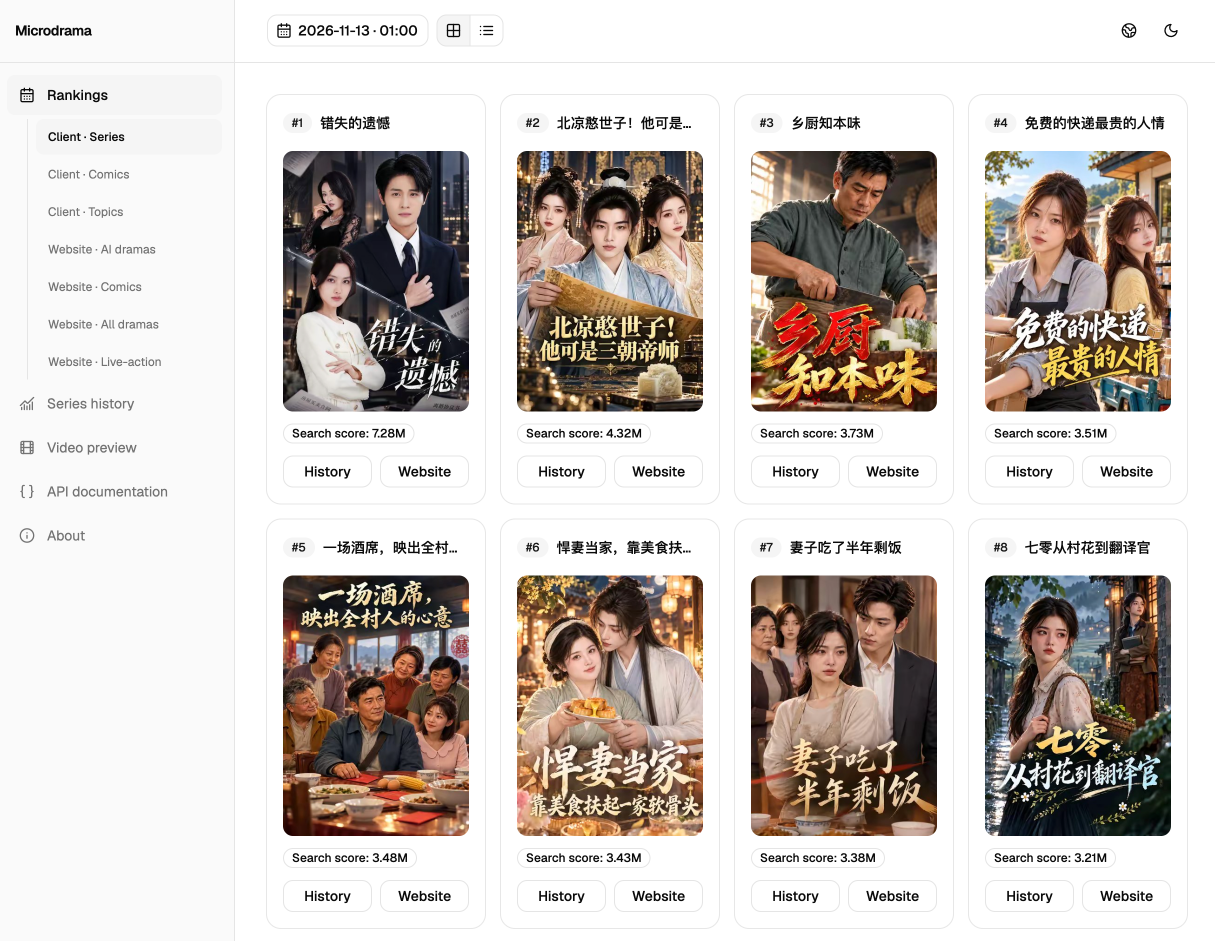}
\caption{Ranking gallery with date selection, series covers, search scores, and links to history and the source website.}
\label{fig:frontend-gallery}
\end{figure}

Figure~\ref{fig:frontend-gallery} shows the web interface for browsing historical snapshots, viewing series metadata, and inspecting rank trajectories.

 \section{Trending Lists}
\label{app:descriptive}

\subsection{Distinct interfaces and content populations}
The web and mobile client interfaces surface distinct series populations with different format and genre profiles. First, cross-interface overlap is small, because although the four web lists contain \NSeries{} distinct series across \NEntries{} entries (Table~\ref{tab:boards}), only \NClientWebOverlap{} of the \NClientSeries{} series recorded on the client trending streams also appear on the web lists. Second, series lengths vary by production mode, as within the web catalog median episode counts reach \EpMedLiveaction{} for live-action, \EpMedAIgenerated{} for AI-generated, and \EpMedAnimated{} for animated lists. Third, genre distributions diverge across lists, with each category emphasizing distinct thematic tags (Figure~\ref{fig:content}B). Because web and mobile trending lists curate separate pools of content, empirical analyses should treat each interface and production category separately rather than pooling them as a single population.

\subsection{Ranking and engagement operate on different timescales}
Web list rankings update on a concentrated daily schedule, whereas audience engagement accumulates continuously across hours. First, rank entry is tied to a daily batch refresh, as \ShareRefreshAtModal{} of the hours that admit new series occur at \RefreshHourUTC{} UTC across \NRefreshDays{} consecutive days, and list order remains static during hours when membership does not change. Second, engagement metrics change nearly every hour, as consecutive series-hour observations register changes in \ShareLikeChange{} of cases for likes and \ShareFavChange{} for favorites, compared with only \ShareHeatChange{} for website heat. Third, tracking a fixed cohort of 19 backward-matched client titles (7 live-action, 7 AI, and 5 animated) shows steady hour-by-hour growth in likes, favorites, and comments across categories (Figure~\ref{fig:dynamics}B). Hourly detail snapshots therefore capture ongoing audience activity that daily ranking updates alone obscure.

\subsection{Displayed prominence is persistent but uneven}
\begin{figure*}[t]
\centering
\includegraphics[width=0.8\textwidth]{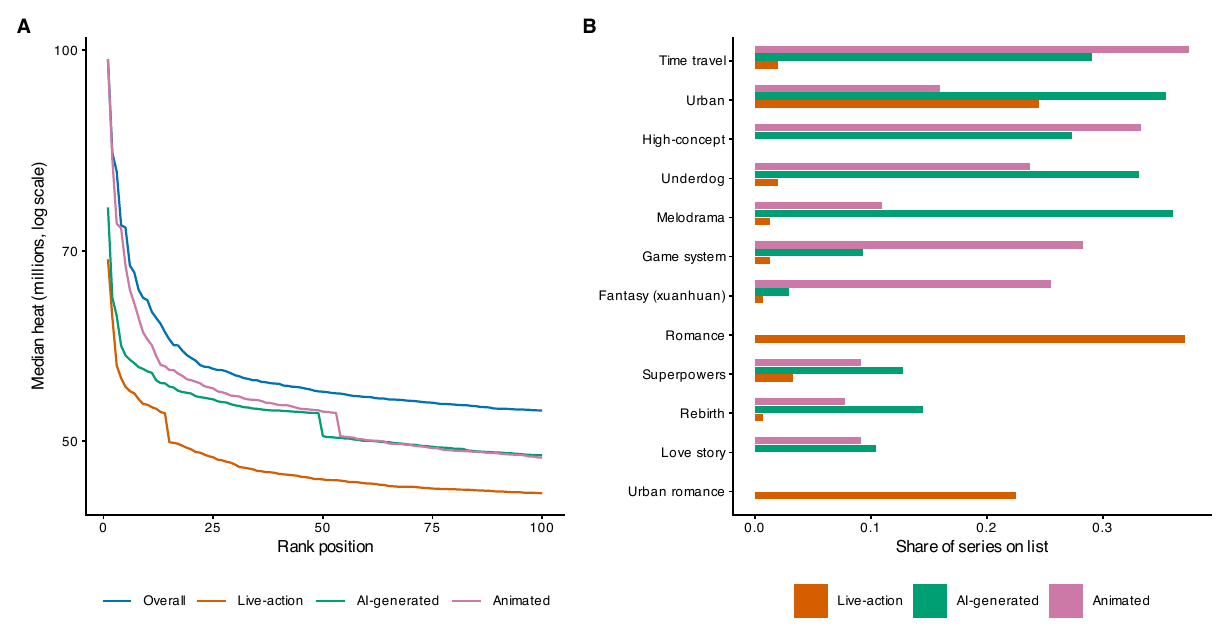}
\caption{Web-list popularity and content. (A) Median heat by rank (log scale). (B) Series shares with each of the 12 most frequent genre tags, translated from Chinese.}
\label{fig:content}
\end{figure*}

\begin{figure*}[t]
\centering
\includegraphics[width=0.8\textwidth]{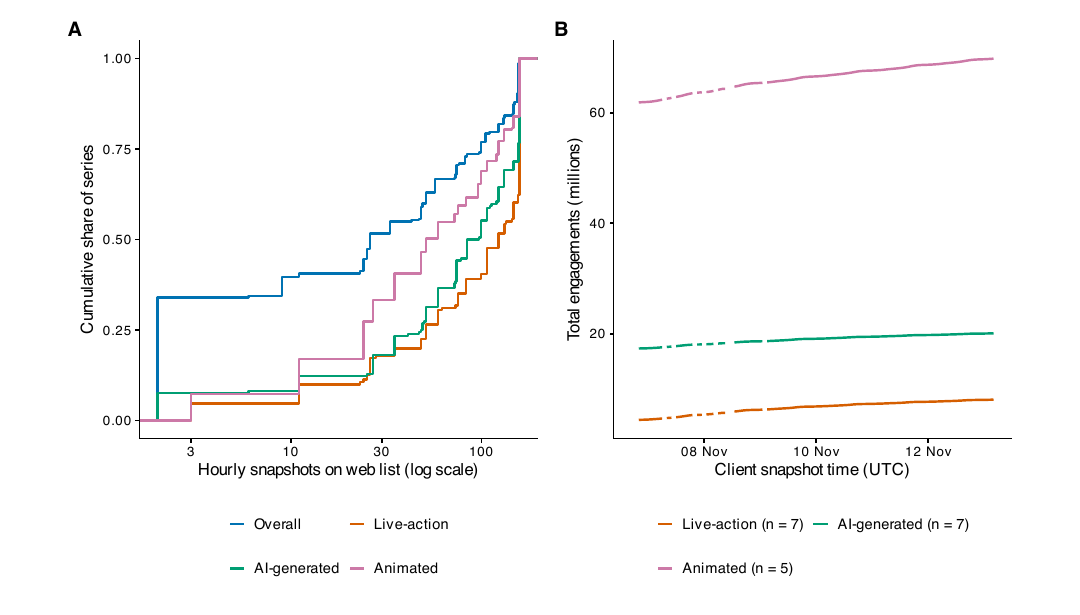}
\caption{Web-list persistence and client-cohort engagement. (A) Cumulative share of series by observed list-hours. (B) Cumulative likes, favorites, and comments at each snapshot for 19 fixed titles, grouped by official category. Gaps interrupt lines, and totals are not new events or unique users.}
\label{fig:dynamics}
\end{figure*}

Series persistence varies sharply between the overall web trending list and category-specific lists, while within-list rank closely tracks heat. First, the overall list turns over much faster than sub-category trending lists, admitting \NNewOverall{} new series per hour on average compared with at most \MaxNewSub{} on the production-mode lists, where the median overall title appears in \MedHoursOverall{} hourly snapshots (Figure~\ref{fig:dynamics}A). Second, cumulative tenure aligns with popularity, as on the overall list total observed list-hours correlate strongly with median heat (Spearman's $\rho=\RhoTenureHeat{}$). Third, displayed rank strictly reflects heat within each snapshot, with at least \MinMono{} of adjacent rank pairs showing non-increasing heat across all web lists (Figure~\ref{fig:content}A). The web trending lists thus maintain a stable prominence hierarchy in which higher-heat series occupy top ranks and accumulate longer total visibility.

 \section{Textual Descriptions}
\label{app:text}
\subsection{Trending-Topic Vocabulary}
\label{app:topicwords}
The client search page includes a trending-topic list (\textit{redian huati bang}) of eight editorial phrases (e.g., ``revenge-drama heroines who forgive no one''). Across \TwRuns{} hourly runs, the module returned \TwAppear{} appearances covering \TwTopics{} distinct topics, with a median persistence of \TwMedianRuns{} runs. We segmented topics using \texttt{jieba} \cite{sun2012jieba} in precise mode with a custom lexicon of genre terms, tropes, and character names, split the genre suffix \textit{ju} (``drama'') from compounds, and dropped single characters and function words. Co-occurrence edges link words appearing in the same phrase, and English plot labels translate the original Chinese tokens.

\begin{figure*}[t]
\centering
\includegraphics[width=0.8\textwidth]{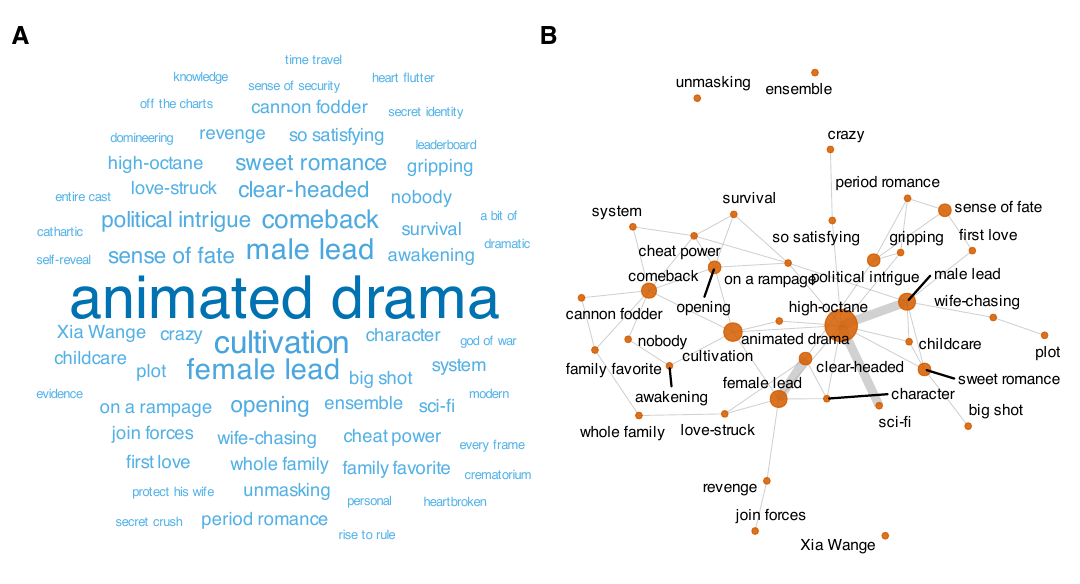}
\caption{Trending-topic vocabulary, deduplicated so that each distinct topic counts once. (A) Word cloud of the 60 most frequent words, sized by frequency. (B) Co-occurrence network of words appearing in at least two topics, where node size shows frequency and edge width the number of topics in which two words co-occur.}
\label{fig:topicdedup}
\end{figure*}

\begin{figure*}[t]
\centering
\includegraphics[width=0.8\textwidth]{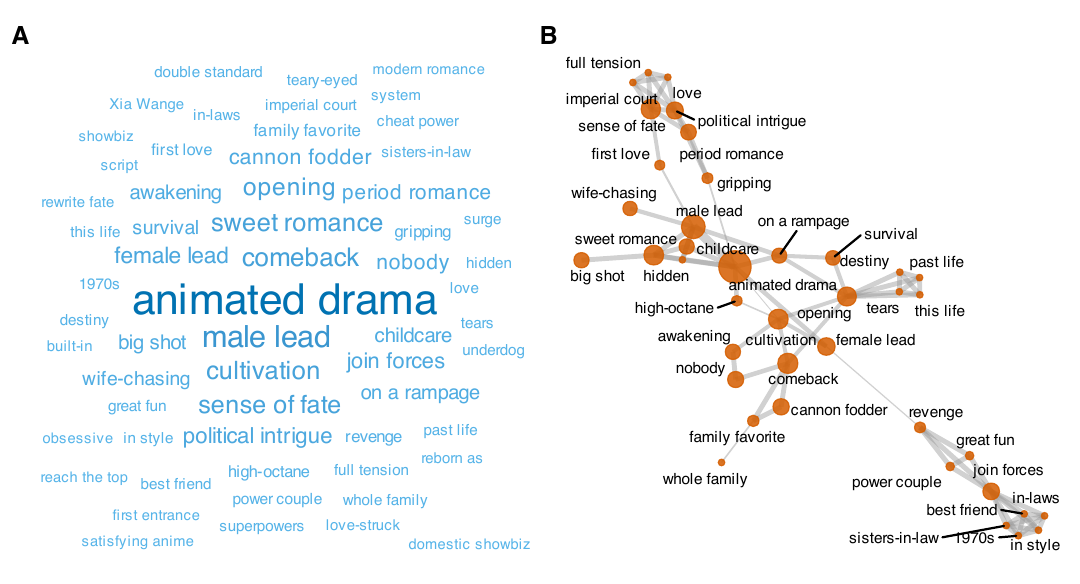}
\caption{Trending-topic vocabulary, weighted by hourly appearances. (A) Word cloud of the 60 most frequent words. (B) Co-occurrence network of the 40 most frequent words, where edge width is the number of hourly appearances in which two words co-occur.}
\label{fig:topicweighted}
\end{figure*}

Editorial trending topics center on animated dramas and recurring genre tropes across both deduplicated and exposure-weighted counts. First, animated content anchors the topic agenda, as \textit{animated drama} (\textit{manju}) is the most frequent term in both the deduplicated corpus (\TwTokens{} tokens and \TwWords{} distinct words across unique topics; \TwTopWordN{} topics; Figure~\ref{fig:topicdedup}) and the hourly appearance-weighted corpus (\TwTokensW{} tokens; \TwTopWordNW{} appearances; Figure~\ref{fig:topicweighted}). Second, secondary terms highlight popular narrative hooks, as \textit{\TwSecond} ranks second in the deduplicated count (\TwSecondN{} topics), whereas \textit{\TwSecondW} and \textit{\TwThirdW} lead the weighted distribution due to longer trending-list retention. Third, word co-occurrence networks cluster around cultivation, status comebacks, fated romance, and court intrigue (Figures~\ref{fig:topicdedup}B and~\ref{fig:topicweighted}B). Platform editors thus use the topic banner to promote genre archetypes and animated series through sustained daily slates.

 \subsection{Title and Synopsis Vocabulary}
\label{app:serieswords}
We applied the same segmentation pipeline (Appendix~\ref{app:topicwords}) to series titles and synopses. Titles cover all \SwTitles{} ranked series (using website detail titles where available and client titles otherwise, with \SwSeasons{} carrying a season marker), while synopses cover the \SwSynopses{} series with website introductions. Co-occurrence edges connect words within the same title or synopsis, evaluated both once per unique series (deduplicated) and across all \SwAppear{} ranked list-hour appearances (weighted).

Series titles highlight dramatic premises and temporal settings, whereas synopses center on family relationships and domestic conflict. First, titles foreground high-concept hooks, as in the deduplicated title corpus (\SwTiTokens{} tokens, \SwTiWords{} distinct words), the top terms are \textit{\SwTiA} (\SwTiAN{} titles), \textit{\SwTiB} (\SwTiBN{}), and \textit{\SwTiC} (\SwTiCN{}; Figure~\ref{fig:titlededup}), while appearance weighting favors \textit{\SwTiAW} (\SwTiAWN{} appearances), \textit{\SwTiBW}, and \textit{\SwTiCW} alongside 1970s/1980s era markers and status reversals (Figure~\ref{fig:titleweighted}). Second, synopses shift focus to interpersonal roles, as across \SwSyTokens{} deduplicated tokens (\SwSyWords{} distinct words), \textit{\SwSyA} (\SwSyAN{} synopses), \textit{\SwSyB} (\SwSyBN{}), and \textit{\SwSyC} (\SwSyCN{}) lead (Figure~\ref{fig:synopsisdedup}), mirrored by \textit{\SwSyAW}, \textit{\SwSyBW}, and \textit{\SwSyCW} in the weighted counts (Figure~\ref{fig:synopsisweighted}). Third, synopsis co-occurrence networks place kinship terms (mother, husband, daughter, and family) at the structural core, connected to hidden identity, awakening, and protection. Promotional text thus operates in two stages, in which titles attract attention with speculative or nostalgic premises, while synopses ground those hooks in familiar kinship dramas.

\newcommand{\swcap}[2]{(A) Word cloud of the 60 most frequent words, sized by frequency. (B) Co-occurrence network of the 40 most frequent words and the 120 strongest links, where node size shows frequency and edge width the number of #1 in which two words co-occur#2.}
\begin{figure*}[t]\centering\includegraphics[width=0.8\textwidth]{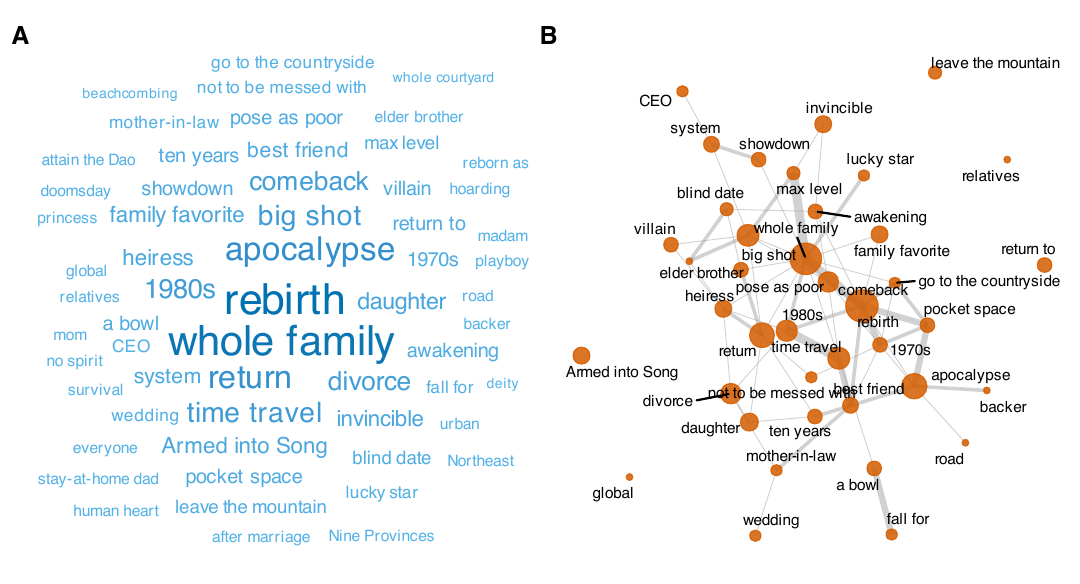}
\caption{Series-title vocabulary, deduplicated so that each series counts once. \swcap{titles}{}}\label{fig:titlededup}\end{figure*}
\begin{figure*}[t]\centering\includegraphics[width=0.8\textwidth]{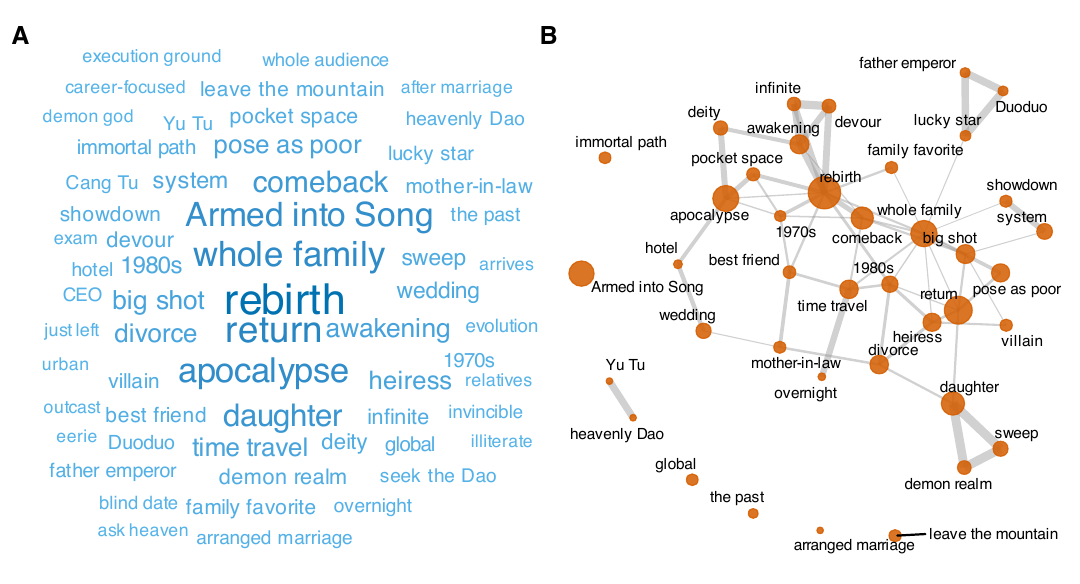}
\caption{Series-title vocabulary, weighted by ranked appearances. \swcap{ranked appearances}{, weighted by appearances}}\label{fig:titleweighted}\end{figure*}
\begin{figure*}[t]\centering\includegraphics[width=0.8\textwidth]{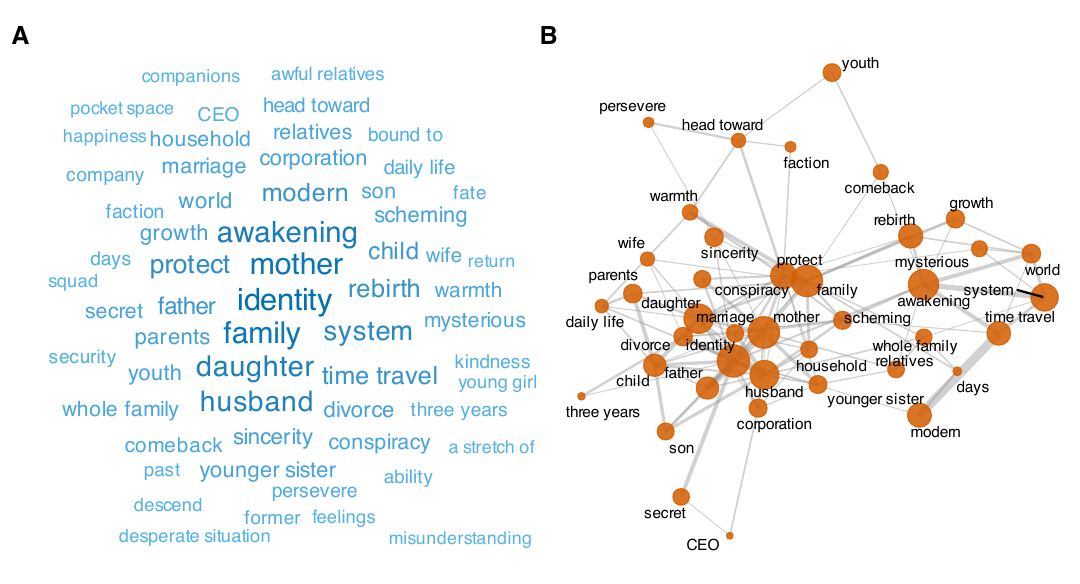}
\caption{Synopsis vocabulary, weighted by ranked appearances. \swcap{ranked appearances}{, weighted by appearances}}\label{fig:synopsisweighted}\end{figure*}
 \subsection{Core-Values Labels: Method and Tests}
\label{app:corevalues}

We classified all \CvN{} website-listed series using TypeSafe Jev (\texttt{\CvModel{}}), a closed-choice decision model. Prompted in Chinese with each series' latest title and synopsis, the model returned a binary decision and candidate probabilities on conformity to China's core socialist values (\CvNLabelled{} valid responses).

Series conforming to core socialist values and non-conforming series receive comparable audience engagement across all website metrics. First, cumulative approval shows only negligible differences, as in Welch tests on $\log(1+x)$ of each series' latest website reading, likes ($\Delta=\CvDiffLike$, 95\% CI [$\CvLoLike$, $\CvHiLike$], $t(\CvDfLike)=\CvTLike$, $p\CvPLike$) and favorites ($\Delta=\CvDiffFav$, CI [$\CvLoFav$, $\CvHiFav$], $t(\CvDfFav)=\CvTFav$, $p\CvPFav$) are only marginally lower among non-conforming titles (Figure~\ref{fig:corevalues}B). Second, active interaction metrics do not differ between groups, including comments ($\Delta=\CvDiffComment$, CI [$\CvLoComment$, $\CvHiComment$], $p\CvPComment$) and shares ($n=\CvNShare$ from 11 November; $\Delta=\CvDiffShare$, CI [$\CvLoShare$, $\CvHiShare$], $p\CvPShare$). Third, overall platform heat is likewise indistinguishable between non-conforming and conforming series ($\Delta=\CvDiffHeat$, CI [$\CvLoHeat$, $\CvHiHeat$], $p\CvPHeat$). Because audience metrics do not reward value conformity, the prevalence of mainstream values on the trending lists likely reflects upstream production and supply rather than viewer selection.
 \section{Covers}
\label{app:covers}
\subsection{Cover Color by Official Category}
\label{app:covercolor}
Cover color distributions vary modestly across official production formats, but the overall category difference is sensitive to specification. First, descriptive color shares differ across formats, as among 72 hash-verified client covers with prior category labels (26 live-action, 24 AI, and 22 animated), animated covers contain more red area (36.1\%) than live-action (19.1\%) or AI covers (25.0\%), and less orange (12.1\% vs.\ 25.8\% and 24.0\%; Figure~\ref{fig:covercolor}A). Second, global omnibus tests show borderline significance, as a Hellinger-distance permutation test across all color bins yields $F(2,69)=1.828$ ($R^2=.050$, $p=.063$). Third, alternative specifications remain fragile, with date-adjusted ($p=.117$) and threshold-varied ($p=.037$ and $.110$) tests straddling conventional significance levels. Production medium alone therefore does not impose a rigid color template on series covers.

By contrast, cover palettes align systematically with narrative setting and genre tags. First, warm tones mark nostalgic settings, as across \CcTagSeries{} tagged website covers tested over \CcTagTests{} tag-color cells (\CcTagSig{} significant at BH-adjusted $q<.05$; Figure~\ref{fig:covercolor}B), rural and period dramas contain significantly more orange ($+\CcRuralOrange$ and $+\CcPeriodOrange$ percentage points; $q \CcRuralOrangeQ$ and $q \CcPeriodOrangeQ$), while period covers contain less black ($\CcPeriodBlack$; $q \CcPeriodBlackQ$). Second, cool and dark tones distinguish modern and speculative worlds, as urban series carry more black ($+\CcUrbanBlack$; $q \CcUrbanBlackQ$), and otherworld series feature more blue and less orange ($+\CcOtherBlue$ and $\CcOtherOrange$). Third, affective genres show distinct contrasts, as melodrama covers contain more red ($+\CcMeloRed$; $q \CcMeloRedQ$), whereas pooled romance titles (\CcRomanceN{} series) use less black ($\CcRomBlack$; $q \CcRomBlackQ$) without a reliable increase in red ($+\CcRomRed$; $q \CcRomRedQ$), and suspense tags are too rare ($\le 2$ series) to evaluate. Designers thus use color primarily to signal spatial and temporal settings rather than broad production categories.

\subsection{Exploratory Cover Composition}
\label{app:covercomposition}

\paragraph{Sample and measurement.}
We analyzed one hash-verified cover for each of \CoverPilotN{} stratified series selected for a transcription pilot (\CoverPilotAIN{} AI-generated, \CoverPilotAnimatedN{} animated, \CoverPilotLiveN{} live-action, and \CoverPilotUnknownN{} unmatched), preferring client cover assets with website covers as fallback. Object detection used YOLO26m (Ultralytics 8.4.212, COCO weights, confidence threshold .25, 960-pixel input), while text detection used Apple Vision OCR (accurate Chinese/English mode, revision 3, language correction disabled). We define text area as the union of OCR bounding boxes divided by image area (coding two text-free covers as zero area) and text location as the union centroid across covers with detected text.

\paragraph{Spatial patterns.}
Pilot covers follow a consistent vertical composition that places prominent title typography in the lower half of the frame. First, text anchors the bottom region, as across the \CoverPilotN{} covers, OCR-detected text boxes concentrate in the lower half of the vertical axis (Figure~\ref{fig:covercomposition}B). Second, text occupies a substantial share of cover area across all production modes, with median union area reaching \CoverPilotAIMedian\% for AI-generated, \CoverPilotAnimatedMedian\% for animated, and \CoverPilotLiveMedian\% for live-action covers (Figure~\ref{fig:covercomposition}C). Third, manual inspection indicates that automated detectors capture standard title blocks reliably but occasionally miss heavily stylized calligraphy or misclassify small decorative elements. Micro-drama covers thus pair upper-frame character imagery with lower-frame title overlays, though fine-grained visual auditing benefits from ground-truth annotation for stylized fonts.
  \section{Audience Engagement}
\label{app:robustness}

\subsection{Heat fixed-effects specifications}
Alternative regression specifications show that website heat reflects persistent cross-series differences far more than short-term fluctuations in individual engagement metrics. First, disaggregating engagement metrics introduces collinearity artifacts, because when entered separately in the daily fixed-effects model (Table~\ref{tab:heatfe}, column 2), likes and favorites take opposite signs because they co-move tightly within series. Second, switching from daily to hourly observations leaves estimates unchanged (column 3), as website heat updates only at the daily refresh and repeats across intervening hours. Third, omitting series fixed effects reverses the sign on likes to positive (\HfeNoSeriesLike{}, 95\% CI [\HfeNoSeriesLikeLo{}, \HfeNoSeriesLikeHi{}]; column 4) by conflating between-series baseline popularity with within-series changes. Modeling platform heat therefore requires series fixed effects and aggregated engagement indices to separate within-title changes from baseline visibility.

\subsection{Correlations among Engagement Metrics}
\begin{figure*}[t]
\centering
\includegraphics[width=.6\textwidth]{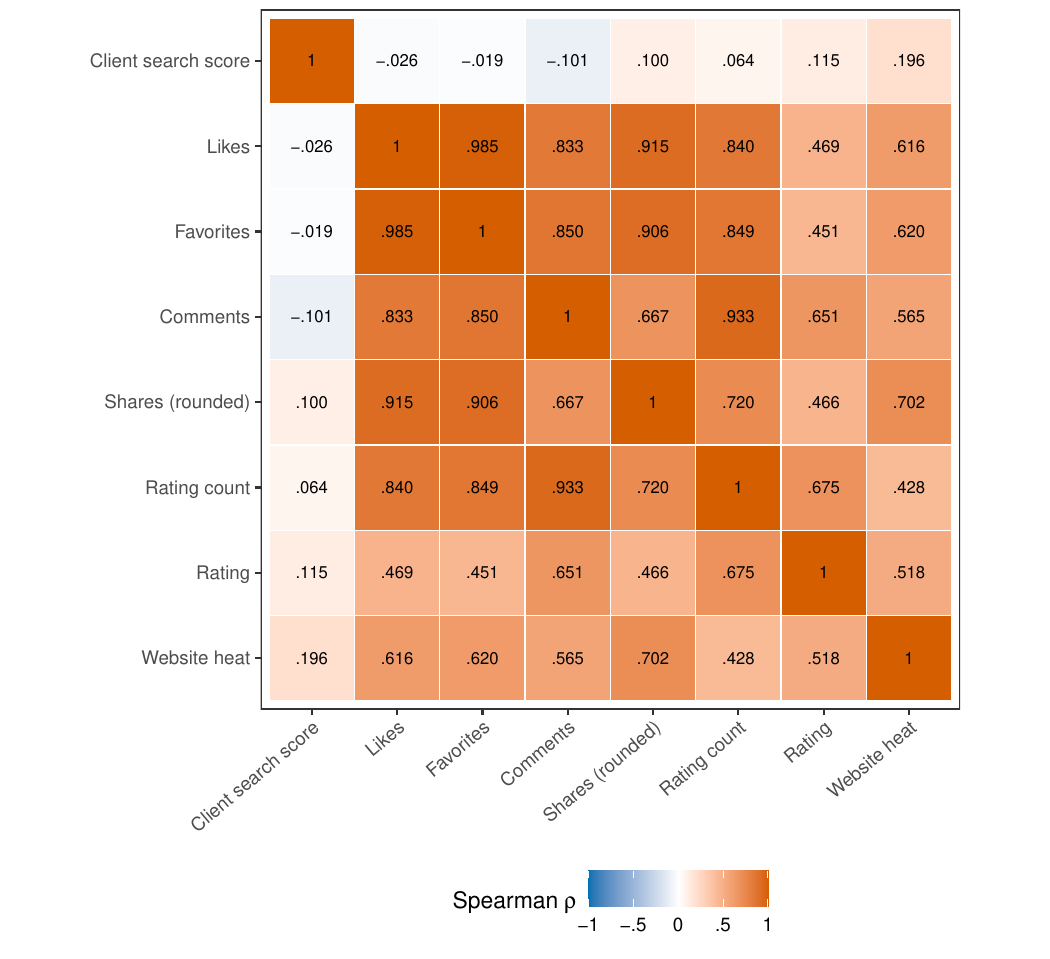}
\caption{Spearman correlations for client short-drama titles and backward-matched website metrics. Cells use available pairs, so sample sizes vary, and shares are approximate. Search score and website heat are distinct measures.}
\label{fig:correlation}
\end{figure*}

Cumulative website engagement metrics correlate strongly with one another but remain decoupled from the mobile client's search score. First, website engagement metrics are highly redundant, as across matched short-drama series likes and favorites correlate at Spearman's $\rho=.985$ across all observations and $\rho=.974$ using per-series medians (Figure~\ref{fig:correlation}). Second, client search scores show near-zero correlation with website likes, favorites, comments, and shares, with coefficients ranging from $-.101$ to $.100$ and all 95\% series-bootstrap intervals spanning zero. Third, pairwise sample sizes vary across cells due to asynchronous cross-platform matching and cumulative growth over a title's lifecycle. Website engagement metrics thus capture a shared dimension of cumulative audience approval, whereas the mobile search score reflects a distinct, interface-specific popularity signal.

 \clearpage
\section{API Request and Response Examples}
\label{app:api}
Below are two example API requests and responses, one retrieving a single entry from the website overall list at a given snapshot and the other fetching metadata for an individual series. Full endpoint specifications are available in the online API documentation.

\paragraph{Snapshot.}

{\footnotesize
\begin{verbatim}
GET /api/v1/snapshots/72126/entries?limit=1
Host: micro-drama.rongxin.me
Accept: application/json
\end{verbatim}
\textit{Response excerpt:}
\begin{verbatim}
{
  "snapshot": {
    "id": "72126",
    "category": "hot-drama",
    "scheduled_at":
    "2026-11-13T06:00:00Z",
    "rank_complete": true
  },
  "items": [
    {
      "rank": 1,
      "series_id": "7702763145030290457",
      "heat": 83874843.0,
      "like_count": 5100000,
      "fetched_at":
    "2026-11-13T06:00:07Z"
    }
  ],
  "page": 1,
  "page_size": 1,
  "total": 100,
  "has_more": true
}
\end{verbatim}
}

\newpage \paragraph{Series.}

{\footnotesize
\begin{verbatim}
GET /api/v1/series/7702763145030290457
Host: micro-drama.rongxin.me
Accept: application/json
\end{verbatim}
\textit{Response excerpt:}
\begin{verbatim}
{
  "series_id": "7702763145030290457",
  "first_seen_at":
    "2026-11-08T02:00:31Z",
  "first_listed_at":
    "2026-11-08T02:00:00Z",
  "updated_at":
    "2026-11-13T06:02:17Z",
  "listed_windows": 250,
  "best_rank": 1,
  "version_fetched_at":
    "2026-11-08T02:00:56Z"
}
\end{verbatim}
} 

\begin{thebibliography}{38}
\providecommand{\natexlab}[1]{#1}

\bibitem[{Abadie(2021)}]{abadie2021using}
Abadie, A. 2021.
\newblock Using Synthetic Controls: Feasibility, Data Requirements, and
  Methodological Aspects.
\newblock \emph{Journal of Economic Literature}, 59(2): 391--425.

\bibitem[{Athey and Imbens(2017)}]{athey2017state}
Athey, S.; and Imbens, G.~W. 2017.
\newblock The {{State}} of {{Applied Econometrics}}: {{Causality}} and {{Policy
  Evaluation}}.
\newblock \emph{Journal of Economic Perspectives}, 31(2): 3--32.

\bibitem[{Athey, Mobius, and Pal(2021)}]{athey2021impact}
Athey, S.; Mobius, M.; and Pal, J. 2021.
\newblock The Impact of Aggregators on Internet News Consumption.
\newblock Technical report, National Bureau of Economic Research.

\bibitem[{Baker et~al.(2025)Baker, Callaway, Cunningham, {Goodman-Bacon}, and
  Sant'Anna}]{baker2025did}
Baker, A.; Callaway, B.; Cunningham, S.; {Goodman-Bacon}, A.; and Sant'Anna, P.
  H.~C. 2025.
\newblock Difference-in-{{Differences Designs}}: {{A Practitioner}}'s
  {{Guide}}.
\newblock arXiv:2503.13323.

\bibitem[{{Caixin}(2024)}]{caixin2024boxoffice}
{Caixin}. 2024.
\newblock Caixin Weekly: Micro Drama Revenues Surpass Box Office as Phenomenal
  Apps Sweep Globally.
\newblock
  https://www.caixinglobal.com/2024-12-06/caixin-weekly-micro-drama-revenues-surpass-box-office-phenomenal-apps-sweep-globally-102265249.html.

\bibitem[{Cao et~al.(2026)Cao, He, Liu, and LC}]{cao2026audience}
Cao, G.; He, T.; Liu, Y.; and LC, {\relax RAY}. 2026.
\newblock Audience in the {{Loop}}: {{Viewer Feedback-Driven Content Creation}}
  in {{Micro}}-Drama {{Production}} on {{Social Media}}.
\newblock In \emph{Proceedings of the 2026 {{CHI Conference}} on {{Human
  Factors}} in {{Computing Systems}}}, {{CHI}} '26, 1--25. New York, NY, USA:
  Association for Computing Machinery.
\newblock ISBN 979-8-4007-2278-3.

\bibitem[{Chen, Valdovinos~Kaye, and Zeng(2021)}]{chen2021positiveenergy}
Chen, X.; Valdovinos~Kaye, D.~B.; and Zeng, J. 2021.
\newblock \#{{PositiveEnergy Douyin}}: Constructing ``Playful Patriotism'' in a
  {{Chinese}} Short-Video Application.
\newblock \emph{Chinese Journal of Communication}, 14(1): 97--117.

\bibitem[{{DataEye}(2025)}]{dataeye2025haiwai}
{DataEye}. 2025.
\newblock {Overseas Micro-Drama Data Report} ({in Chinese}).
\newblock https://www.dataeye.com/\allowbreak report.html?\allowbreak
  key=\allowbreak
  DataEye2025\allowbreak\%E5\allowbreak\%B9\allowbreak\%B4\allowbreak\%E6\allowbreak\%B5\allowbreak\%B7\allowbreak\%E5\allowbreak\%A4\allowbreak\%96\allowbreak\%E5\allowbreak\%BE\allowbreak\%AE\allowbreak\%E7\allowbreak\%9F\allowbreak\%AD\allowbreak\%E5\allowbreak\%89\allowbreak\%A7\allowbreak\%E6\allowbreak\%95\allowbreak\%B0\allowbreak\%E6\allowbreak\%8D\allowbreak\%AE\allowbreak\%E6\allowbreak\%8A\allowbreak\%A5\allowbreak\%E5\allowbreak\%91\allowbreak\%8A.

\bibitem[{Fei(1992)}]{fei1992soil}
Fei, X. 1992.
\newblock \emph{From the {{Soil}}: {{The Foundations}} of {{Chinese Society}}}.
\newblock University of California Press, 1 edition.
\newblock ISBN 978-0-520-07795-9.

\bibitem[{Gillespie(2014)}]{gillespie2014relevance}
Gillespie, T. 2014.
\newblock The Relevance of Algorithms.
\newblock In \emph{Media Technologies: {{Essays}} on Communication,
  Materiality, and Society}, volume 167, 167. Cambridge, MA.

\bibitem[{Goncalves and Ng(2026)}]{goncalves2026global}
Goncalves, A.; and Ng, Y. M.~M. 2026.
\newblock Global {{YouTube Trending Dataset}} (2022--2025): {{Three Years}} of
  {{Platform-Curated}}, {{Cross-National Trends}} in {{Digital Culture}}.
\newblock \emph{Proceedings of the International AAAI Conference on Web and
  Social Media}, 20(1): 2817--2827.

\bibitem[{Guess et~al.(2023)Guess, Malhotra, Pan, Barber{\'a}, Allcott, Brown,
  {Crespo-Tenorio}, Dimmery, Freelon, Gentzkow, {Gonz{\'a}lez-Bail{\'o}n},
  Kennedy, Kim, Lazer, Moehler, Nyhan, Rivera, Settle, Thomas, Thorson,
  Tromble, Wilkins, Wojcieszak, Xiong, {\noopsort{jonge}}{de Jonge}, Franco,
  Mason, Stroud, and Tucker}]{guess2023feeds}
Guess, A.~M.; Malhotra, N.; Pan, J.; Barber{\'a}, P.; Allcott, H.; Brown, T.;
  {Crespo-Tenorio}, A.; Dimmery, D.; Freelon, D.; Gentzkow, M.;
  {Gonz{\'a}lez-Bail{\'o}n}, S.; Kennedy, E.; Kim, Y.~M.; Lazer, D.; Moehler,
  D.; Nyhan, B.; Rivera, C.~V.; Settle, J.; Thomas, D.~R.; Thorson, E.;
  Tromble, R.; Wilkins, A.; Wojcieszak, M.; Xiong, B.; {\noopsort{jonge}}{de
  Jonge}, C.~K.; Franco, A.; Mason, W.; Stroud, N.~J.; and Tucker, J.~A. 2023.
\newblock How Do Social Media Feed Algorithms Affect Attitudes and Behavior in
  an Election Campaign?
\newblock \emph{Science}, 381(6656): 398--404.

\bibitem[{Guo, Li, and Yang(2025)}]{guo2025civilizing}
Guo, Y.; Li, Y.; and Yang, T. 2025.
\newblock Civilizing Social Media: {{The}} Effect of Geolocation on the
  Incivility of News Comments.
\newblock \emph{New Media \& Society}, 27(5): 2996--3016.

\bibitem[{Hanney(2026)}]{hanney2026new}
Hanney, R. 2026.
\newblock New Directions in Vertical Storytelling: Creative Shifts in
  Micro-Drama for Mobile Audiences.
\newblock \emph{Media Practice and Education}, 0(0): 1--19.

\bibitem[{Huszár et~al.(2022)Huszár, Ktena, O'Brien, Belli, Schlaikjer, and
  Hardt}]{huszar2022amplification}
Huszár, F.; Ktena, S.~I.; O'Brien, C.; Belli, L.; Schlaikjer, A.; and Hardt,
  M. 2022.
\newblock Algorithmic amplification of politics on {Twitter}.
\newblock \emph{Proceedings of the National Academy of Sciences}, 119(1):
  e2025334119.

\bibitem[{{International Trade Administration}(2025)}]{ita2025microdrama}
{International Trade Administration}. 2025.
\newblock China {{Entertainment Micro}}-Drama {{Industry}}.
\newblock
  https://www.trade.gov/market-intelligence/china-entertainment-micro-drama-industry.

\bibitem[{Kayal et~al.(2025)Kayal, Mettes, Dehmamy, and
  Park}]{kayal2025popularity}
Kayal, P.; Mettes, P.; Dehmamy, N.; and Park, M. 2025.
\newblock Large {{Language Models Are Natural Video Popularity Predictors}}.
\newblock In Che, W.; Nabende, J.; Shutova, E.; and Pilehvar, M.~T., eds.,
  \emph{Findings of the {{Association}} for {{Computational Linguistics}}:
  {{ACL}} 2025}, 11432--11464. Vienna, Austria: Association for Computational
  Linguistics.
\newblock ISBN 979-8-89176-256-5.

\bibitem[{Liu(2026)}]{liu2026producing}
Liu, C. 2026.
\newblock Producing {{Commissioned Drama}} for {{Chinese Short Video
  Platforms}}: {{Power}}, {{Dependence}} and {{Creative Control}}.
\newblock \emph{Television \& New Media}, 15274764261461453.

\bibitem[{Liu et~al.(2026{\natexlab{a}})Liu, Li, Peng, Wang, Liao, Chang, Gao,
  Zhao, Liu, Jiang, and Niu}]{liu2026bridging}
Liu, Q.; Li, J.; Peng, Z.; Wang, S.; Liao, Z.; Chang, S.; Gao, B.; Zhao, H.;
  Liu, M.; Jiang, J.; and Niu, L. 2026{\natexlab{a}}.
\newblock Bridging Visual Dynamics and Narrative Reasoning: Multimodal Large
  Language Models for Short Drama Quality Assessment.
\newblock In \emph{Proceedings of the ACM Web Conference 2026}, WWW '26,
  7890--7901. New York, NY, USA: Association for Computing Machinery.
\newblock ISBN 9798400723070.

\bibitem[{Liu et~al.(2026{\natexlab{b}})Liu, Lu, Peng, Qian, Peng, and
  Shen}]{liu2026seeing}
Liu, X.; Lu, Y.; Peng, Q.; Qian, S.; Peng, Y.; and Shen, C. 2026{\natexlab{b}}.
\newblock Seeing the {{Surreal}}: {{Mapping Surrealism}} in {{Photorealistic
  AI-Generated Images Using Large Language Models}}.
\newblock \emph{Computational Communication Research}, 8(2).

\bibitem[{Liu et~al.(2026{\natexlab{c}})Liu, Zhao, Li, Wang, and
  Chen}]{liu2026unveiling}
Liu, X.; Zhao, J.; Li, Z.; Wang, D.; and Chen, A. 2026{\natexlab{c}}.
\newblock Unveiling the {{Making}} of {{Trending Topics}} on a {{Digital
  Platform}}: {{A Research Note}} on {{Chinese Sina Weibo}}.
\newblock \emph{Social Science Computer Review}, 44(2): 259--270.

\bibitem[{Lu et~al.(2025)Lu, Pan, Xu, and Xu}]{lu2025decentralized}
Lu, Y.; Pan, J.; Xu, X.; and Xu, Y. 2025.
\newblock Decentralized Propaganda in the Era of Digital Media: {{The}} Massive
  Presence of the {{Chinese}} State on {{Douyin}}.
\newblock \emph{American Journal of Political Science}.

\bibitem[{Mukerjee and Shen(2025)}]{mukerjee2025antecedents}
Mukerjee, S.; and Shen, C. 2025.
\newblock The {{Antecedents}} and {{Manifestations}} of {{Political
  Polarization}} in {{Visual Media}}: {{Key Questions}} and {{Future
  Directions}}.
\newblock \emph{Political Communication}, 42(1): 208--214.

\bibitem[{Nieborg and Poell(2018)}]{nieborg2018platformization}
Nieborg, D.~B.; and Poell, T. 2018.
\newblock The Platformization of Cultural Production: {{Theorizing}} the
  Contingent Cultural Commodity.
\newblock \emph{New Media \& Society}, 20(11): 4275--4292.

\bibitem[{Peng, Lock, and Ali~Salah(2024)}]{peng2024automated}
Peng, Y.; Lock, I.; and Ali~Salah, A. 2024.
\newblock Automated Visual Analysis for the Study of Social Media Effects:
  Opportunities, Approaches, and Challenges.
\newblock \emph{Communication Methods and Measures}, 18(2): 163--185.

\bibitem[{Qin et~al.(2026)Qin, Chen, Yu, Cheng, Cheng, Luo, and
  Wei}]{qin2026sparse}
Qin, Y.; Chen, S.-Z.; Yu, Z.; Cheng, S.; Cheng, T.; Luo, J.; and Wei, Z. 2026.
\newblock Beyond {{Sparse Rewards}}: {{A New Benchmark}} and {{Structure-Aware
  Graph Alignment}} for {{Micro-Drama Understanding}}.
\newblock In Lee, H.-Y.; Dev, S.; Li, J.~J.; and Elliott, D., eds.,
  \emph{Proceedings of the 2026 {{Conference}} on {{Empirical Methods}} in
  {{Natural Language Processing}}}, 31686--31711. Budapest, Hungary:
  Association for Computational Linguistics.
\newblock ISBN 979-8-89176-459-0.

\bibitem[{{QuestMobile}(2026)}]{questmobile2026autumn}
{QuestMobile}. 2026.
\newblock {China} Mobile Internet Autumn Report ({in Chinese}).

\bibitem[{{RustFS}(2023)}]{rustfs}
{RustFS}. 2023.
\newblock {RustFS}: An Open-Source, {S3}-Compatible High-Performance Object
  Storage System.
\newblock https://github.com/rustfs/rustfs.

\bibitem[{Sharma and Peng(2024)}]{sharma2024visual}
Sharma, M.; and Peng, Y. 2024.
\newblock How Visual Aesthetics and Calorie Density Predict Food Image
  Popularity on Instagram: A Computer Vision Analysis.
\newblock \emph{Health Communication}, 39(3): 577--591.

\bibitem[{Sun(2012)}]{sun2012jieba}
Sun, A. 2012.
\newblock Jieba: {Chinese} Text Segmentation.
\newblock https://github.com/fxsjy/jieba.

\bibitem[{Tang and Wang(2025)}]{tang2025reelshort}
Tang, W.; and Wang, Y. 2025.
\newblock {{ReelShort}} as a {{New Template}} of {{International Short}}-Drama
  {{Business}}: {{Platformisation}}, {{Glocalisation}}, and {{De-Westernised
  Practices}}.
\newblock \emph{Global Media and China}, 20594364251366414.

\bibitem[{Urman, Hann{\'a}k, and Baumann(2026)}]{urman2026googletrends}
Urman, A.; Hann{\'a}k, A.; and Baumann, J. 2026.
\newblock {{GoogleTrendArchive}}: {{A Year-Long Archive}} of {{Real-Time Web
  Search Trends Worldwide}}.
\newblock \emph{Proceedings of the International AAAI Conference on Web and
  Social Media}, 20(1): 2936--2948.

\bibitem[{Wang et~al.(2025)Wang, Li, Wei, {Songxueyu}, Song, {Xiaxiaoqiang},
  Zeng, Chen, {Liuliu}, Xu, and Xu}]{wang2025dramaad}
Wang, X.; Li, X.; Wei, Y.; {Songxueyu}; Song, Y.; {Xiaxiaoqiang}; Zeng, F.;
  Chen, Z.; {Liuliu}; Xu, G.; and Xu, T. 2025.
\newblock From Long Videos to Engaging Clips: A Human-Inspired Video Editing
  Framework with Multimodal Narrative Understanding.
\newblock In \emph{Proceedings of the 2025 Conference on Empirical Methods in
  Natural Language Processing: Industry Track}, 2764--2781. Association for
  Computational Linguistics.

\bibitem[{Xue et~al.(2026)Xue, Zhang, Wang, Kim, and Song}]{xue2026sensation}
Xue, H.; Zhang, J.; Wang, X.; Kim, D.; and Song, Y. 2026.
\newblock A {{Computational Model}} of {{Message Sensation Value}} in {{Short
  Video Multimodal Features}} That {{Predicts Sensory}} and {{Behavioral
  Engagement}}.
\newblock \emph{Computational Communication Research}, 8(1): 1--27.

\bibitem[{Yang(2026)}]{openhongguo}
Yang, B. 2026.
\newblock Openhongguo.
\newblock https://github.com/beilunyang/openhongguo.

\bibitem[{Yang and Peng(2020)}]{yang2020importance}
Yang, T.; and Peng, Y. 2020.
\newblock The {{Importance}} of {{Trending Topics}} in the {{Gatekeeping}} of
  {{Social Media News Engagement}}: {{A Natural Experiment}} on {{Weibo}}.
\newblock \emph{Communication Research}, 49(7): 994--1015.

\bibitem[{Yu and Margolin(2024)}]{yu2024topic}
Yu, C.; and Margolin, D. 2024.
\newblock Topic Territoriality and the Cost of Civility: Examining the Impact
  of {{IP}} Address Disclosure on {{Weibo}}.
\newblock \emph{Journal of Computer-Mediated Communication}, 29(6): zmae018.

\bibitem[{Zhong et~al.(2026)Zhong, Xie, Bi, Yang, and Chen}]{zhong2026storytr}
Zhong, X.; Xie, Y.; Bi, G.; Yang, J.; and Chen, G. 2026.
\newblock {{StoryTR}}: {{Narrative-Centric Video Temporal Retrieval}} with
  {{Theory}} of {{Mind Reasoning}}.
\newblock arXiv:2604.23198.
\newblock arXiv:2604.23198.

\end{thebibliography}
\end{document}